\documentclass[superscriptaddress,amsmath,amssymb,twocolumn,pra,floatfix]{revtex4-1}
\usepackage{amssymb,amsmath}
\usepackage{graphicx}
\usepackage{enumitem}
\usepackage[dvipsnames,usenames]{xcolor}
\usepackage[breaklinks=true,colorlinks=true,linkcolor=blue,urlcolor=blue,citecolor=blue]{hyperref}

\begin{document}

\title{Heat, particle and chiral currents in a boundary driven bosonic ladders in presence of gauge fields} 

\author{Bo Xing}
\affiliation{Science,  Mathematics  and  Technology  Cluster, Singapore University of Technology and Design, 8 Somapah Road, 487372 Singapore}
\author{Xiansong Xu}
\affiliation{Science,  Mathematics  and  Technology  Cluster, Singapore University of Technology and Design, 8 Somapah Road, 487372 Singapore}
\author{Vinitha Balachandran}
\affiliation{Science,  Mathematics  and  Technology  Cluster, Singapore University of Technology and Design, 8 Somapah Road, 487372 Singapore}
\author{Dario Poletti}
\affiliation{Science,  Mathematics  and  Technology  Cluster, Singapore University of Technology and Design, 8 Somapah Road, 487372 Singapore}

\date{\today}     

\begin{abstract}
Quantum systems can undergo phase transitions and show distinct features in different phases. The corresponding transport properties can also vary significantly due to the underlying quantum phase. We investigate the transport behaviour of a two-legged bosonic ladder in a uniform gauge field, which is known to have a Meissner-like phase and a vortex phase in the absence of dissipation.  The ladder is coupled to bosonic baths at different temperatures, and we study it using the non-equilibrium Green's function method. In particular, we show the presence of a chiral current and how it is affected by the temperature bias and the dissipation strength. We also demonstrate that the opening of a gap between the lower and upper energy band results in the possibility of tuning heat and particle transport through the ladder. We show that for system parameters for which the ground state is in a vortex phase, the system is more sensitive to external perturbations.
\end{abstract}

\maketitle

\section{Introduction}\label{sec:intro} 
Low-dimensional systems can present interesting transport properties such as ballistic, superdiffusive, diffusive, subdiffusive, and insulating \cite{BertiniZnidaric2020, Prosen2011b, Znidaric2011} behaviours. An interesting research direction is to tune the transport properties by varying external parameters so that systems can be better suited for applications such as energy conversion \cite{BenentiWhitney2017}. One possibility to strongly change the transport through a system is to rely on the underlying presence of phase transitions, whether in the ground state \cite{Sachdev2011}, or out of equilibrium \cite{DiehlZoller2008, DiehlZoller2010, DallaTorreAltman2010, LudwigMarquardt2013, CarusottoCiuti2013, BaumannEsslinger2010}.      
In this work, we analyse in detail the transport properties of a low-dimensional system coupled to two bosonic baths as we tune the system and bath's parameters. In particular, we consider a ladder of non-interacting bosons which can undergo a quantum phase transition by varying the magnitude of the gauge field. More precisely, the system can be in a Meissner phase, in which the ground state is characterized by a current that flows on the boundary of the system and not inside it, or in a vortex phase, in which the ground state presents vortices of currents. The ground-state properties of bosonic ladder with a gauge field have been studied thoroughly also in the presence of interactions \cite{Kardar1986, Granato1990, DennistonTang1995, Nishiyama2000, OrignacGiamarchi2001, DonohueGiamarchi2001, DharParamekanti2012, DharParamekanti2013, PetrescuLeHur2013, ChaShin2011, CrepinSimon2011, TovmasyanHuber2013, PetrescuLeHur2013, WeiMueller2014, TokunoGeorges2014, PiraudSchollwock2015, GreschnerVekua2015, BuserSchollwock2019, KamarGiamarchi2019}. 
Its ground-state properties have been studied experimentally first with Josephson junction arrays \cite{FaziovanderZant2001, vanOudenaardenMooij1996}, and thanks to the possibility of generating synthetic gauge fields \cite{{DalibardOhberg2011, GoldmanSpielman2014}}, also in ultracold atoms \cite{AtalaBloch2014, AnGadway2017}. We note that synthetic gauge fields can potentially be generated also in trapped ions experiments \cite{BermudezPorras2011}.  

However, the non-equilibrium properties of bosonic ladders with gauge fields driven by bosonic baths at their edges are far less explored. In the presence of local ``density'' baths, it was shown that transport is not only affected by the quantum phase transition, but also by the possibility of a gap opening between the two energy bands \cite{GuoPoletti2016}. In Ref. \cite{RivasMartin-Delgado2017} the authors considered larger two-dimensional geometries and the presence of defects. The interplay of the gauge field and interactions under boundary driving has, so far, only been considered in Ref. \cite{GuoPoletti2017} in which the authors analysed hard-core bosons and studied the transport properties as a function of the average density. However, the above studies were performed within the framework of Gorini-Kossakowski-Sudarshan-Lindblad master equations \cite{GoriniSudarshan1976, Lindblad1976} which have shown to have limitations when analysing currents within systems \cite{WichterichMichel2007, ThingnaHanggi2013, PurkayasthaKulkarni2016, XuWang2017, HoferBrunner2017}. We thus focus here on using the non-equilibrium Green's functions formalism \cite{CaroliSaint-James1971, MeirWingreen1992, HaugJauho2008, ProciukDunietz2010, Aeberhard2011, ZimbovskayaPederson2011, NikolicThygesen2012, DharHanggi2012, WangThingna2014,  Ryndyk2016} which is known to be able to represent exactly the non-equilibrium properties of non-interacting systems, even in presence of strong system-environment coupling.       

In this work, we perform a thorough analysis of the transport properties by considering the four possible emerging scenarios: whether the ground state has a Meissner or vortex phase, and whether the energy spectrum is gapped or not. The Green's function approach allows us to study exactly the current flowing through each bond, and this allows us to recognize the reminiscence of the current patterns typical of the Meissner or vortex phases and how the system transits from one to the other. 
We also consider the overall particle and heat currents through the system, unveiling which regions of the parameter space result in larger currents as we vary the temperatures in the baths and the strength of the coupling to them. By analysing the effect of strong system-bath coupling, we show that the parameter space regions for which the ground state is in the Meissner phase are more robust against the dissipation from the system-environment coupling. 

The paper is organized as follows: in Sec. \ref{sec:model} we introduce the set-up, summarize the key aspects of the energy band structure of the bosonic ladder, and present the non-equilibrium Green's function method we use. Our analysis of chiral currents is presented in \ref{sec:Chiral}, while particle and heat transport properties are detailed in \ref{sec:Transport}. Finally, we draw our conclusions in Sec. \ref{sec:conclusions}

\section{Model and Methods}\label{sec:model}
\subsection{Two-legged Bosonic ladder}
\begin{figure}[htp]
    \centering
    \includegraphics[width=\columnwidth]{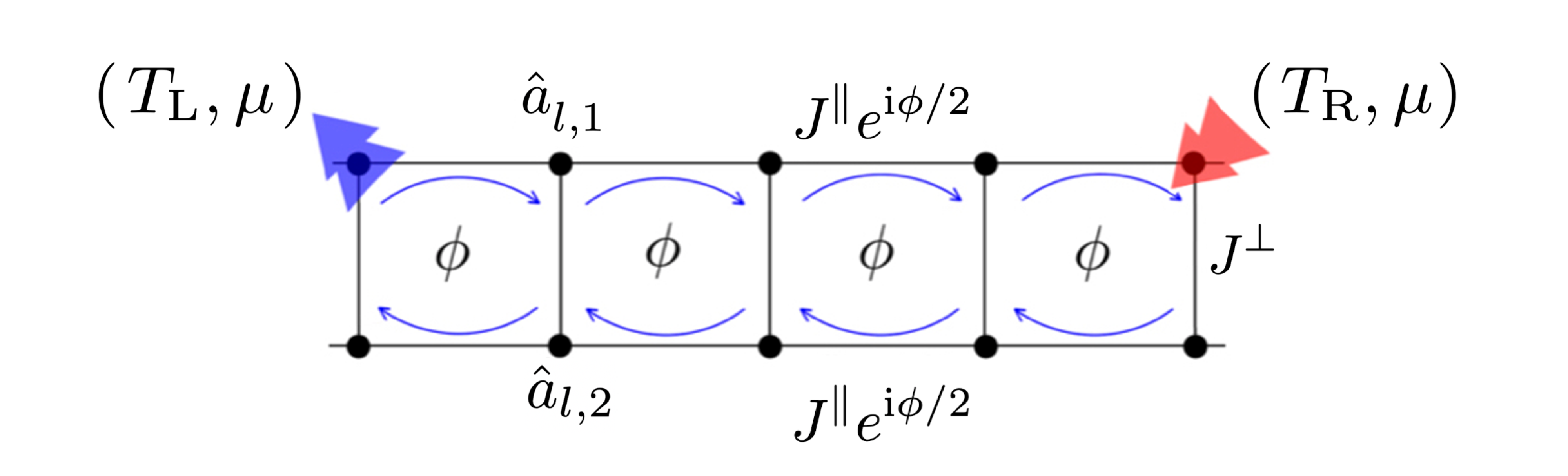}
    \caption{A schematic representation of our set-up. The ladder consists of two coupled legs, with local bosonic excitations described by $\hat{a}_{l,p}$ ($\hat{a}_{l,p}^{\dagger}$), the annihilation (creation) operators at rung $l$, where $p=1,2$ refers to the top or bottom leg.
    $J^{\perp}$ and  $J_{{\|}}$ are the tunnelling amplitude on the rungs of the ladder and along the legs respectively.
  A gauge field imposes a phase factor $\phi$ when hopping around a plaquette. 
  The coupling to the left and right bath are represented by the blue and red double-arrow.
  Each bath is characterized by temperatures $T_{\mathrm{L}}$ or $T_{\mathrm{R}}$ and a common chemical potential $\mu$.
  The length of the ladder considered throughout the article is $L=32$.
}
    \label{fig:model}
\end{figure}

The system we study is a two-legged ladder of non-interacting bosons subjected to a uniform magnetic field with Hamiltonian,
\begin{equation}\label{eqn:Hs}
    \begin{aligned}
      \hat{H}_{\rm S} = &-\left( J^{\|} \sum_{l,p} e^{i\left( -1 \right)^{p+1}\phi/2} \; \hat{a}_{l,p}^{\dagger}\hat{a}_{l+1,p} \right. \\
                      &\left. + J^{\perp}\sum_{l} \hat{a}_{l,1}^{\dagger}\hat{a}_{l,2} + \text{H.c} \right) + V\sum_{l,p}\hat{a}_{l,p}^{\dagger}\hat{a}_{l,p}
    ,\end{aligned}
\end{equation}
where $\hat{a}_{l,p}$ ($\hat{a}_{l,p}^{\dagger}$) is the bosonic annihilation (creation) operator at the $l$-th rung and $p$-th leg of the ladder, $J^{\|}$ and $J^{\perp}$ are the
tunnelling amplitude along the legs and rungs respectively. In the presence of a gauge field, the bosons acquire a phase $\phi$ by tunnelling around a plaquette, with a sign that depends on the direction of the field circulation as shown in Fig. \ref{fig:model}. $V$ is the local potential of the system. Throughout the article, we consider a ladder with a length $L=32$ and a local potential $V/J^{\parallel}=8$ \footnote{Simulations for longer systems, even $L=64$ and different local potentials $V/J^{\parallel}$ have also been performed and are consistent with the results here obtained.}.

\subsection{Energy structure of two-legged bosonic ladder} \label{sec:Band}
\begin{figure}[htp]
    \centering
    \includegraphics[width=\columnwidth]{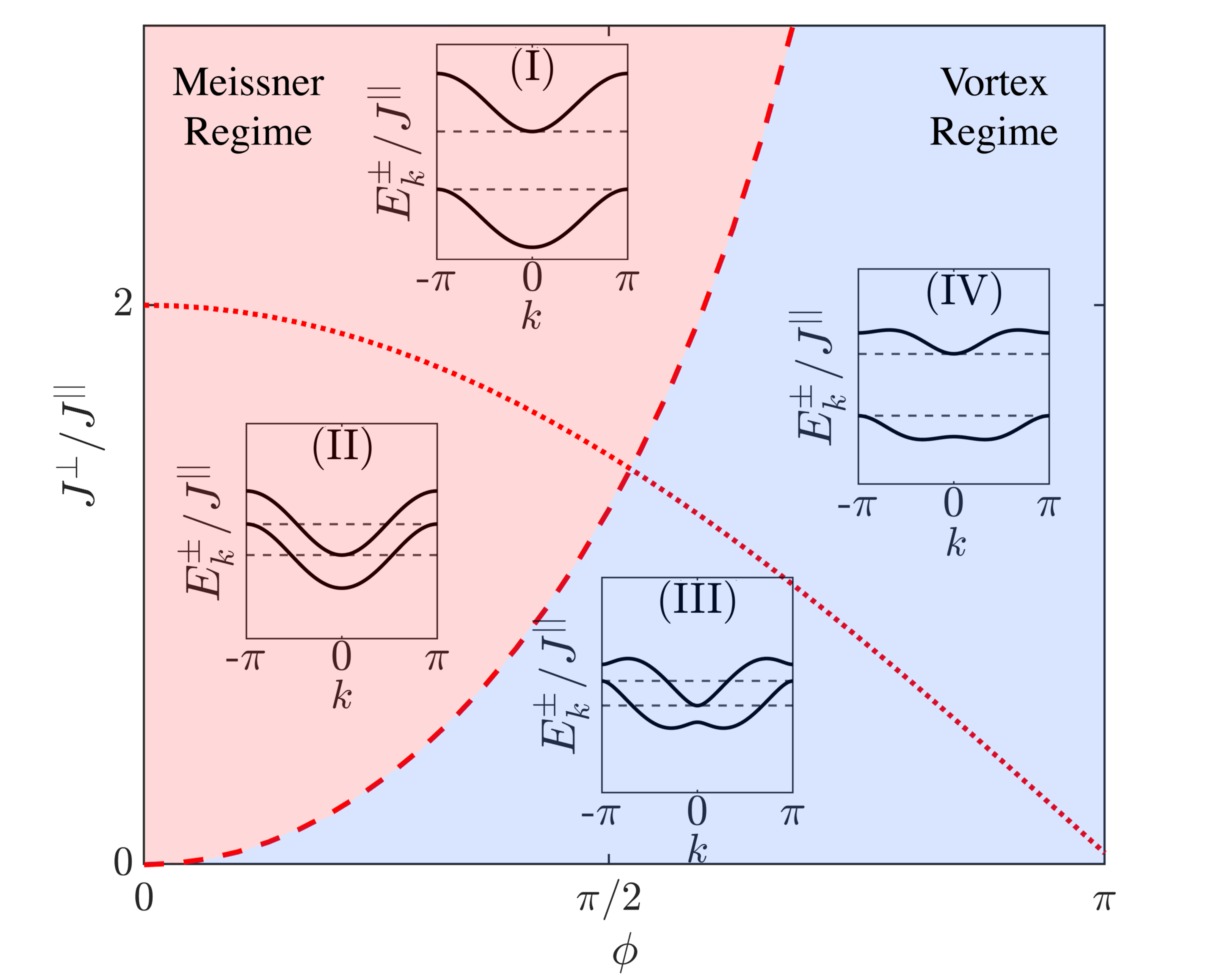}
    \caption{Energy band structures of a two-legged bosonic ladder as a function of  $J^{\perp}/J^{\|}$ and $\phi$.
    The red dotted line corresponds to Eq. (\ref{eqn:line1}) and the dashed line to Eq. (\ref{eqn:line2}). 
    Together they divide the parameter space into four distinct regions (I) to (IV).  Each region is characterized by a qualitatively different energy band structure.
    In each subplot, we show the energy band structure, i.e., $E^{\pm}_{k}/J^{\|}$ from Eq. (\ref{eq:quaienergies}) versus the quasi momentum $k$. The dashed lines in the subplots indicate the energy levels of $\mathrm{max}( E^{-}_{k}/J^{\|} )$ and $\mathrm{min}( E^{+}_{k}/J^{\|} )$. 
    The Meissner to vortex quantum phase transition occurs across the red dashed line corresponding to Eq. (\ref{eqn:line2}).
}
    \label{fig:espec}
\end{figure} 

The single-particle Hamiltonian in Eq. (\ref{eqn:Hs}) with periodic boundary condition can be diagonalized using the Bogoliubov transformation and has a two-band structure \cite{Kardar1986},
\begin{equation}
    \hat{H}_{\rm S}= \sum_{k}E_{k}^{\pm}\hat{\alpha}_{k,\pm}^{\dagger}\hat{\alpha}_{k,\pm}
,\end{equation}
with eigenenergies 
\begin{align}
    E_{k}^{\pm} = & V - 2 J^{\|}\cos\left( \phi/2 \right) \cos\left( k \right) \nonumber\\
                  &\pm \sqrt{J^{\perp \ 2} + \left[ 2J^{\|}\sin\left( \phi/2 \right) \sin\left( k \right)  \right]^2 }. \label{eq:quaienergies}
\end{align}
Here, $\hat{\alpha}_{k,\pm}$ ($\hat{\alpha}^{\dagger}_{k,\pm}$) is the annihilation (creation) operator  of the quasi particle at momentum $k$ in the upper ($+$) or lower ($-$) band. Depending on the choice of $J^{\perp}/J^{\|}$ and $\phi$, the energy spectrum of the ladder can be classified into four typical scenarios, as shown in Fig. \ref{fig:espec}.
\begin{enumerate}[label=(\Roman*)]
  \item Energy bands have a single minimum at $k=0$, and an energy gap separates the two bands. This is seen when $J^{\perp}/J^{\|}$ is large and the $\phi$ is small.
  \item Energy bands have a single minimum at $k=0$, and no energy gap separates the two bands. This is typical for small values of $J^{\perp}/J^{\|}$ and $\phi$.
  \item Lower band is double-well shaped with two minima, and no energy gap separates the two bands. This is seen in the limit of very large $\phi$.
  \item Lower band is double-well shaped with two minima, and an energy gap separates the two bands. This is seen in the limit of very large values of $J^{\perp}/J^{\|}$ and $\phi$.
\end{enumerate}

The opening of the energy gap occurs at the critical value for the perpendicular tunnelling $J_{c1}^{\perp} $ 
\begin{equation}\label{eqn:line1}
    J_{c1}^{\perp} = 2J^{\|} \cos{\left( \phi/2 \right)}
,\end{equation}
and is denoted by dotted lines in Fig. \ref{fig:espec} whereas the degeneracy of ground state occurs at a tunnelling amplitude $J_{c2}^{\perp}$ given by
\begin{equation}\label{eqn:line2}
    J_{c2}^{\perp} = 2J^{\|}\tan{\left( \phi/ 2\right)}\sin{\left( \phi/2 \right)}
\end{equation}
and is denoted by the dashed lines. For $ J^{\perp}< J_{c2}^{\perp}$ , the ground state of the system is in the Meissner phase, in which a particle current only occurs along the edges of the ladder with no inner rung current. This parameter space is denoted in Fig. \ref{fig:espec} in red. With $ J^{\perp}>J_{c2}^{\perp}$, the ladders enters a vortex phase with finite inner rung current. This parameter space is denoted in Fig. \ref{fig:espec} in blue. The focus of our paper is to study how these phases affect the transport properties of the system. We also show that the presence of energy gap between the two bands can be used to alter transport properties. 
In the following, we work in units for which $J^{\|}=k_B=\hbar=1$. 

\subsection{Non-equilibrium set-up}
The ladder is coupled to two bosonic baths at different temperatures at its edges as shown in Fig. \ref{fig:model}. The baths are modelled by a collection of non-interacting bosons with Hamiltonian,
\begin{align}
    \hat{H}_{\rm L/R} =& \sum_{k}E_{k,{\rm L/R}} \; \hat{b}_{k,{\rm L/R}}^{\dagger}\hat{b}_{k,{\rm L/R}}.
\end{align}
Here, $\hat{b}_{k,{\rm L/R}}$ ($\hat{b}^{\dagger}_{k,{\rm L/R}}$) is the annihilation (creation) operator for a bosonic excitation with energy $E_{k,{\rm L/R}}$ in the left ($\mathrm{L}$) or right ($\mathrm{R}$) bath. The baths are assumed to be at thermal equilibrium characterized by the Bose-Einstein distribution at temperature $T_{\rm L/R}$. In our studies we keep the left bath at a fixed temperature $T_{\mathrm{L}}=0.1$, which is low enough so that the currents will be affected by the properties of the system near the ground state. We also fix the bath chemical potentials $\mu$ such that the ground-state occupation is 
\begin{equation}
    \bar{n}_{0}\left( T \right) =\frac{1}{e^{\left( E_{0}-\mu \right) / T}-1}=\frac{1}{e^{\Delta/ T}-1}
\end{equation}
where $\Delta = E_0 -\mu = 0.1$ for all system parameters $J^{\perp}$ and $\phi$ for a given temperature $T$. $E_{0}$ is the ground-state energy of the system. By fixing $\Delta$, the occupation of the excited states solely depends on the temperature and energy difference between the excited state and the ground state. 
In this way, we can gain a clearer understanding of the roles of temperature and energy band structure in the non-equilibrium ladder.  

The system-bath coupling is defined by the Hamiltonian,
\begin{align}
  \hat{H}_{I,{\rm L/R}} =& \sum_{k}c_{k,{\rm L/R}}\left( \hat{a}_{{\rm L/R}}^{\dagger}\hat{b}_{k,{\rm L/R}} + \hat{b}_{k,{\rm L/R}}^{\dagger}\hat{a}_{{\rm L/R}} \right) \label{eqn:sbint}
,\end{align}
where $c_{k,\mathrm{L}/\mathrm{R}}$ denotes the strength of the coupling and $\hat{a}_{\rm L/R},\;\hat{a}^\dagger_{\rm L/R}$ are the system operators at the edges of the ladder as indicated by Fig. \ref{fig:model}. Note that the total number of bosons within the system and bath is conserved for this particular choice of system-environment interaction.  

\subsection{Green's function formalism} 
To study the transport in our system, we use the non-equilibrium Green's function formalism \cite{CaroliSaint-James1971, MeirWingreen1992, HaugJauho2008, ProciukDunietz2010, Aeberhard2011, ZimbovskayaPederson2011, NikolicThygesen2012, DharHanggi2012, WangThingna2014,  Ryndyk2016} which we briefly describe in this section. The central object of this formalism is the retarded and advanced Green's function  $G^{\rm r,a}(E)$
\begin{equation}
  G^{\rm r,a}(E)=\frac{1}{E-\hat{H}_{\rm S}-\Sigma_{\mathrm{L}}^{\rm r,a}(E)-\Sigma_{\mathrm{R}}^{\rm r,a}(E)}
,\end{equation}
where $\Sigma^{\rm r, a}_{\mathrm{L}/\mathrm{R}}\left( \omega \right)$ are the self-energy terms that model the effects of the baths on the isolated system and can be written in terms of the free Green's function of the baths $g^{\rm r,a}_{\mathrm{L}/\mathrm{R}}=(E \pm i\epsilon- \hat{H}_{\mathrm{L}/\mathrm{R}})^{-1}$ and the coupling Hamiltonian $\hat{H}_{I,\mathrm{L}/\mathrm{R}}$,
\begin{equation}\label{eqn:selfenergy}
  \Sigma_{\mathrm{L}/\mathrm{R}}^{\rm r,a}\left( E \right) = \hat{H}_{I,\mathrm{L}/\mathrm{R}}g_{\mathrm{L}/\mathrm{R}}^{\rm r,a}\left( E \right)\hat{H}_{I,\mathrm{L}/\mathrm{R}}^{\dagger}
.\end{equation}

The bath spectral density, also known as the level-width function, can be defined as,
\begin{align}
    \Gamma_{\mathrm{L}/\mathrm{R}}\left( E \right) & = {\rm i}(\Sigma^{\rm r}_{\mathrm{L}/\mathrm{R}} - \Sigma^{\rm a}_{\mathrm{L}/\mathrm{R}}) \nonumber \\
                                               & = 2\pi \sum_{k} |c_{k,\mathrm{L}/\mathrm{R}}|^{2}\delta\left( E - E_{k,\mathrm{L}/\mathrm{R}} \right)
,\end{align}
which characterizes the coupling between the system and baths.

In the following we consider baths with Ohmic spectral density $\Gamma_{\mathrm{L}/\mathrm{R}}\left( E \right) = \gamma E $, where $\gamma$ is the effective system-bath coupling strength for each bath \cite{DittrichZwerger1998}.    

We can thus write the steady-state single-particle density matrix $\rho$ as  
\begin{align}\label{eqn:corrm}
    &\rho_{m,n}=\left\langle \hat{a}_{m}^{\dagger} \hat{a}_{n} \right\rangle \nonumber \\ 
    &= \frac{1}{2\pi} \int_{-\infty}^{\infty} \!\!\! d E \left\{ \left[ G^{\rm r}\!\left( E \right) \Gamma_{\mathrm{L}}\left( E \right) G^{\rm a}\!\left( E \right) \right]_{n,m} f\left( E, T_{\mathrm{L}}, \mu \right)\right. \nonumber \\
                        &\left.  +\left[ G^{\rm r}\!\left( E \right) \Gamma_{\mathrm{R}}\left( E \right) G^{\rm a}\!\left( E \right) \right]_{n,m} f\left( E,T_{\mathrm{R}},\mu \right)\right\},  
\end{align} 
where $f(E,T,\mu) = 1/({e^{\left( E-\mu \right)/{T}} - 1})$ is the Bose-Einstein distribution function and $\mu$ is the chemical potential of the baths, which we take to be identical for both baths. 

It follows that the particle current $\mathcal{J}_{P}$ and heat current $\mathcal{J}_{Q}$ are given by Landauer formula \cite{Landauer1957, Landauer1970}
\begin{equation}\label{eqn:pmeirwingreen}
    \mathcal{J}_{P} = \frac{1}{2\pi}\int_{-\infty}^{\infty} \!\!\! d E \; \mathcal{T}\!\left( E \right) \left[ f{\left( E,T_{\mathrm{L}},\mu \right)} - f{\left( E,T_{\mathrm{R}},\mu \right)} \right],
\end{equation}
and 
\begin{equation}\label{eqn:hmeirwingreen}
    \mathcal{J}_{Q} = \frac{1}{2\pi}\int_{-\infty}^{\infty} \!\!\! d E \left( E - \mu \right) \mathcal{T}\!\left( E \right) \left[ f{\left( E,T_{\mathrm{L}},\mu \right)}- f{\left( E,T_{\mathrm{R}},\mu \right)} \right], 
\end{equation}
where $\mathcal{T}(E) = \mathrm{Tr}\left[ G^{\rm r}\left( E \right)\Gamma_{\mathrm{L}}\left( E \right)G^{\rm a}\left( E \right)\Gamma_{\mathrm{R}}\left( E \right)\right]$ is the transmission function \cite{CaroliSaint-James1971}. 
It is important to note that Eq. (\ref{eqn:pmeirwingreen}) and Eq. (\ref{eqn:hmeirwingreen}) are valid for two-terminal devices even when a magnetic field is present \cite{Datta1995}.

\section{Chiral Current properties} \label{sec:Chiral}
In this section, we study the steady-state chiral current of the ladder driven by two bosonic baths. The chiral current $\mathcal{J}_{c}$ is defined as
\begin{equation}
    \mathcal{J}_{c} = \sum_{l}\left( \mathcal{J}_{l,1}-\mathcal{J}_{l,2} \right)/L
,\end{equation}
 where $L$ is the length of the ladder, $\mathcal{J}_{l,1}$ and $\mathcal{J}_{l,2}$ are the local particle currents in the upper and lower legs of the ladder. The local currents $\mathcal{J}_{l, p}$ can be obtained from the continuity equations
\begin{equation}\label{eqn:cont1}
    \frac{\partial \langle \hat{n}_{l,1} \rangle }{\partial t} = \mathcal{J}_{l-1,1} - \mathcal{J}_{l,1} - \mathcal{J}_{l,1\to 2}
,\end{equation}
\begin{equation}\label{eqn:cont2}
    \frac{\partial \langle \hat{n}_{l,2} \rangle }{\partial t} = \mathcal{J}_{l-1,2} - \mathcal{J}_{l,2} - \mathcal{J}_{l,2\to 1}
.\end{equation}
Here, $\mathcal{J}_{l,1\to2}$ is the current in the $l$-th rung of the ladder and can be computed as
\begin{equation}
    \mathcal{J}_{l, 1 \rightarrow 2}=\left\langle\mathrm{i} J^{\perp} \hat{a}_{l, 1}^{\dagger} \hat{a}_{l, 2}+\mathrm{H.c.}\right\rangle
,\end{equation}
and
\begin{equation}
  \mathcal{J}_{l, p}=\left\langle\mathrm{i} J^{ \|}e^{\mathrm{i}(-1)^{p+1} \phi / 2} \hat{a}_{l, 1}^{\dagger} \hat{a}_{l+1,1}+\mathrm{H.c.}\right\rangle
.\end{equation}

\subsection{Steady-state chiral current}\label{sec:ChiralA}
\begin{figure}[htp]
    \centering
    \includegraphics[width=\columnwidth]{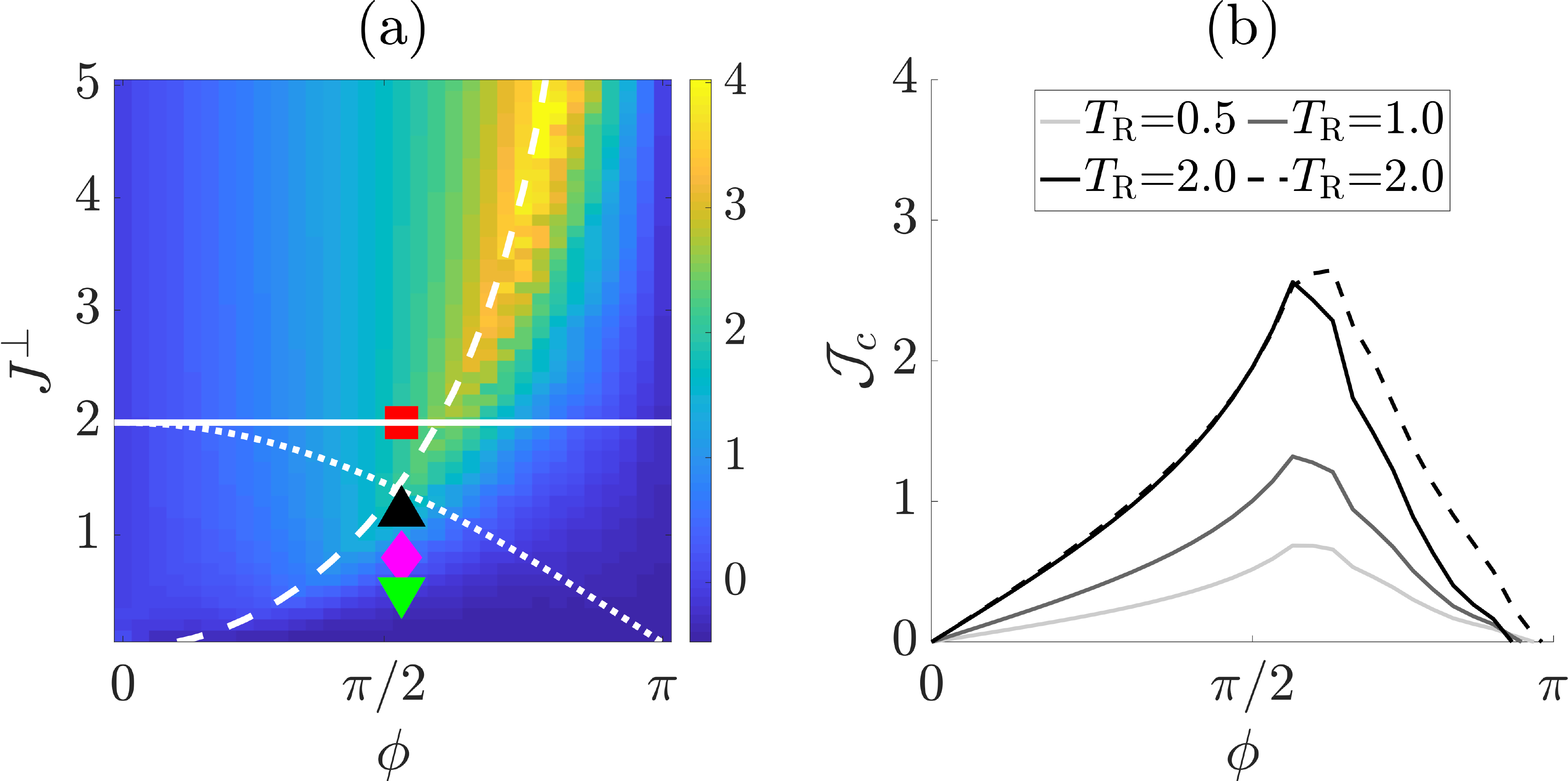}
    \caption{(a) The chiral current $\mathcal{J}_{c}$ as a function $J^{\perp}$ and $\phi$. 
    The white dotted and dashed lines are identical to the red lines in Fig. \ref{fig:espec}.
    The solid white line indicates the location of a horizontal cut at $J^{\perp}=2.0$. 
    The four markers specify the location of the different system parameters used in Fig. \ref{fig:loccurr} and Fig. \ref{fig:loccurr2}.
    (b) $\mathcal{J}_{c}$ of the horizontal cuts at various $T_{\mathrm{R}}$ and $\gamma$.
    The solid lines correspond to coupling strengths $\gamma=0.1$ and dashed lines correspond to  $\gamma=0.5$. The left and right bath potential is $\mu=E_{0}-\Delta$ with $\Delta=0.1$ and the length of the ladder is $L=32$. The other bath parameters are $\gamma=0.1$, $T_{\mathrm{L}}=0.1$ and $T_{\mathrm{R}}=2.0$.
    }
    \label{fig:chiral}
\end{figure} 

In Fig. \ref{fig:chiral}(a), we plot the chiral current $\mathcal{J}_{c}$ as a function of $J^{\perp}$ and $\phi$.  The system-bath couplings are chosen to be $\gamma=0.1$ whereas the temperatures of the left and right bath are kept at $T_{\mathrm{L}}=0.1$ and $T_{\mathrm{R}}=2.0$ respectively. Fig. \ref{fig:chiral}(a) shows that the chiral current is maximum around the dashed line which marks the transition from a single minima band structure to a double minima band. This is similar to the chiral current behaviour for the ground state. 

The dependence on temperature bias and system-bath coupling is studied in Fig. \ref{fig:chiral}(b). Interestingly, the chiral current increases with the increase in temperature bias, while its dependence on system-bath coupling strength is influenced by the energy band structure of the system. 
In particular, we observe that the chiral current is much more robust against changes in the coupling strength when the underlying ground state is in the Meissner phase. For the region where the underlying ground state is in the vortex phase, the chiral current can change significantly with the coupling strength. 

\subsection{Reminiscence of Meissner and vortex phases}
\begin{figure}[htp]
    \centering
    \includegraphics[width=\columnwidth]{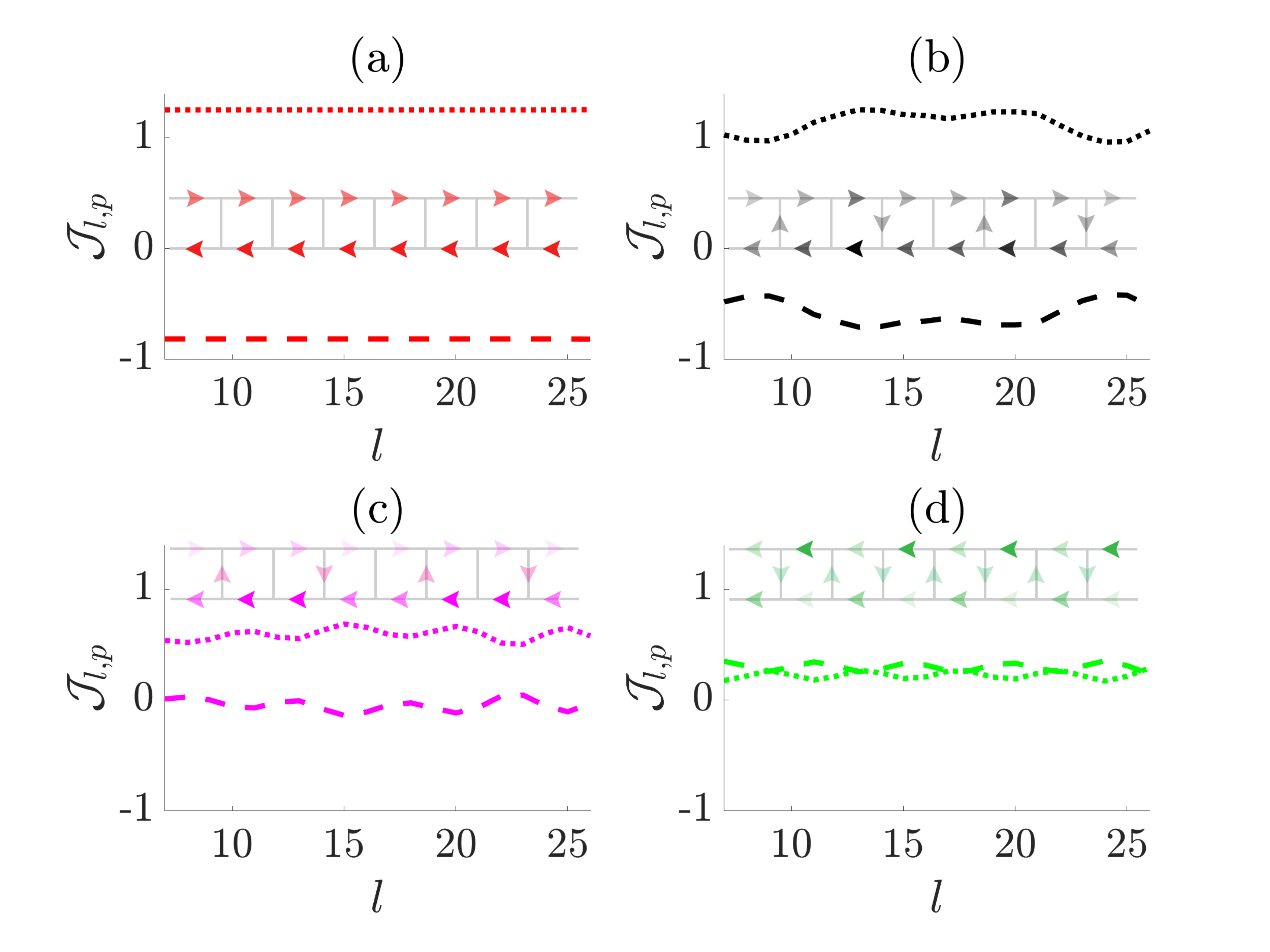}
    \caption{Local particle current, $\mathcal{J}_{l,p}$, along the 20 most central sites of a $L=32$ ladder, where $l$ indicates the index of the site along the leg. The system parameters for (a-d) are $\phi=1.6$, $J^{\perp}= 2.0,1.2,0.8,0.5$ respectively.
    These parameters correspond to the four markers in Fig. \ref{fig:chiral}(a).
    Among the four sets of parameters, (a) is located before the Meissner to vortex transition and (b-d) are located after the transition.
    The dashed and dotted lines represent $\mathcal{J}_{l,p}$ along the upper and lower legs respectively.
    The ladder inset offers a visualization of $\mathcal{J}_{l,p}$ pattern, where the arrow points in the direction of $\mathcal{J}_{l,p}$ and the intensity of colour represents the strength of $\mathcal{J}_{l,p}$. The bath parameters are $\gamma=0.1$, $T_{\mathrm{L}}=0.1, T_{\mathrm{R}}=2$ and $\mu=E_{0}-\Delta$ with $\Delta=0.1$. }
    \label{fig:loccurr}
\end{figure}

\begin{figure}[htp]
    \centering
    \includegraphics[width=\columnwidth]{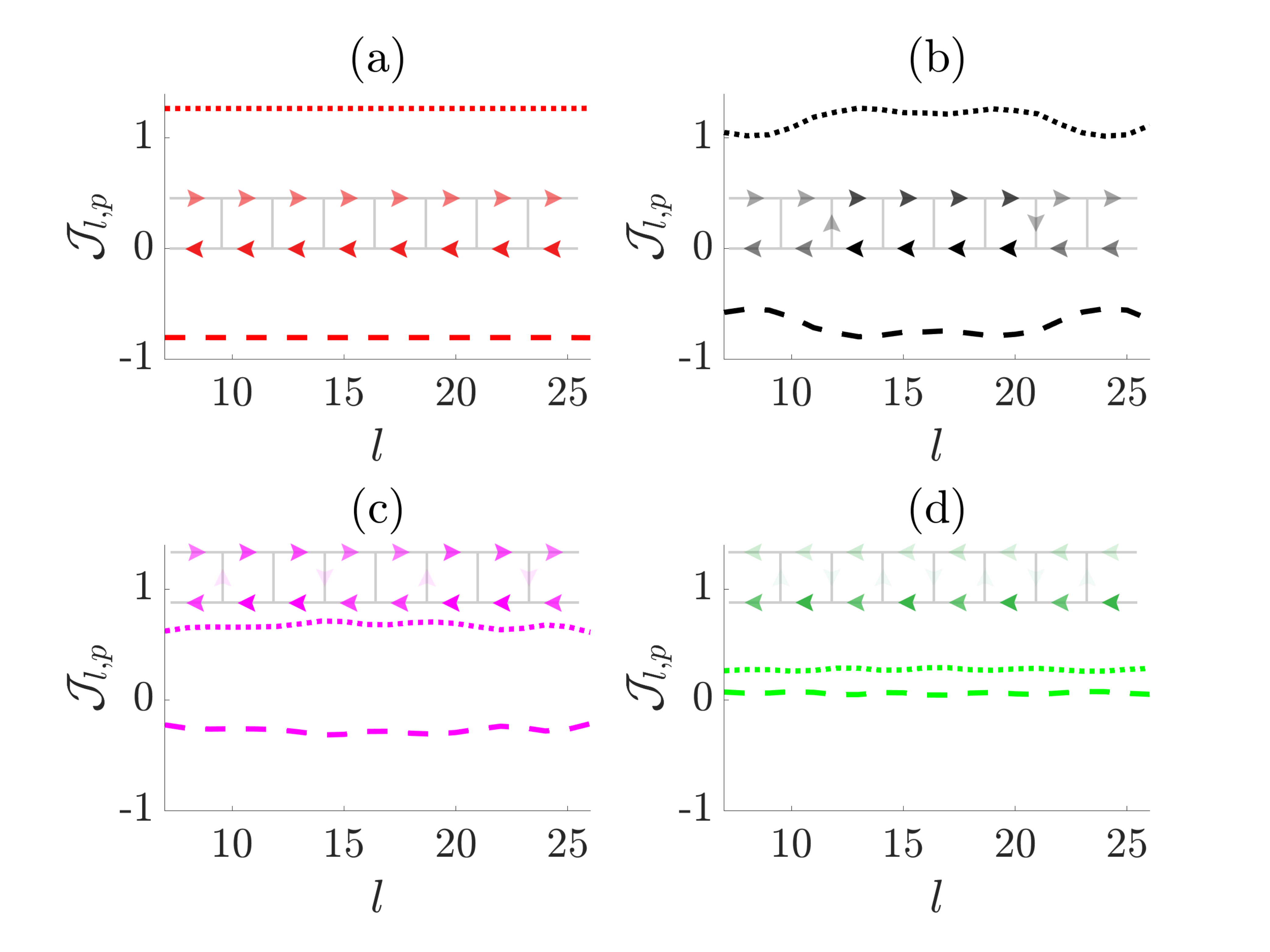}
    \caption{Identical analysis as for Fig. \ref{fig:loccurr}, except with a stronger system-bath coupling $\gamma=0.5$.}
    \label{fig:loccurr2}
\end{figure}
As discussed in Sec. \ref{sec:Band}, by manipulating the tunnelling amplitude along the rungs $J^{\perp}$ and the phase $\phi$, a quantum phase transition between the Meissner and vortex phases occurs in the ground state of the non-interacting bosonic ladder. 
As shown in Fig. \ref{fig:chiral}, even in this non-equilibrium scenario we can observe signatures of the underlying phase transition. 
To probe further the reminiscence of Meissner/vortex phases in the driven ladder, we study the local particle current, $\mathcal{J}_{l,p}$, along the $20$ most central sites of the ladder in Figs. \ref{fig:loccurr} and \ref{fig:loccurr2}. 
We use the term ``forward current'' to denote the current flowing along with the temperature bias (from right to left) and  ``backward current'' to denote the current flowing opposite to the temperature bias. In Fig. \ref{fig:loccurr} we consider $\gamma=0.1$, $T_{\mathrm{L}}=0.1$, $T_{\mathrm{R}}=2$, $\mu=E_{0}-\Delta$ with $\Delta=0.1$. For parameters at which the ground state of the ladder is in the Meissner phase, we find strong forward and backward currents in different legs of the ladder with little current modulation. One such scenario is depicted in Fig. \ref{fig:loccurr}(a). Panels (b)-(d) in Fig. \ref{fig:loccurr} show modulations in the current in both legs, a characteristic of the vortex phase. Note that these panels correspond to parameters in the vortex phase in the ground-state phase diagram. In addition, we find a weaker backward current in the legs that completely disappears in Fig. \ref{fig:loccurr}(d) for which the current $\mathcal{J}_{l,p}$ in both legs are in the forward direction. Note that each panel of Fig. \ref{fig:loccurr} has a schematic inset depicting the intensity and direction of the current (current flows in the direction of the arrows and the magnitude increases with the intensity of the colour). 

In Fig. \ref{fig:loccurr2}, we consider the case of stronger system-bath coupling $\gamma=0.5$. Our results show that the structure of the local current can be very different when the coupling is increased. The local current structure in the parameter space for which the ground state is in the Meissner phase, as shown in Fig. \ref{fig:loccurr2}(a), does not change significantly with the coupling strength. On the other hand, the current modulations in Fig. \ref{fig:loccurr2}(b-d) are suppressed and a stronger backward current is found in the parameter space for which the ground state is in the vortex regime. These observations suggest that the parameter space for which the ground state is in the vortex phase is more sensitive to changes in $\gamma$. The behaviour of the local currents is in agreement with our findings in Sec. \ref{sec:ChiralA}, in which we already observed that, compared to the parameter space for which the ground state is in the vortex phase, the current structure is more robust against changes in the coupling strength in the parameter space for which the ground state is in the Meissner phase.

\section{Transport Across the System} \label{sec:Transport}
In this section, we study the total particle current $\mathcal{J}_{P}$ and heat current $\mathcal{J}_{Q}$ across the ladder. We show that depending upon the energy band structure of the ladder and system-bath coupling, we see a different response of the particle and heat currents. We will consider two different system-bath coupling strengths $\gamma=0.1$ and $\gamma=0.5$ respectively in Secs. \ref{ssec:g01} and \ref{ssec:g05}.       

\subsection{System-bath coupling strength $\gamma=0.1$}\label{ssec:g01}
\begin{figure}[htp]
    \centering
    \includegraphics[width=\columnwidth]{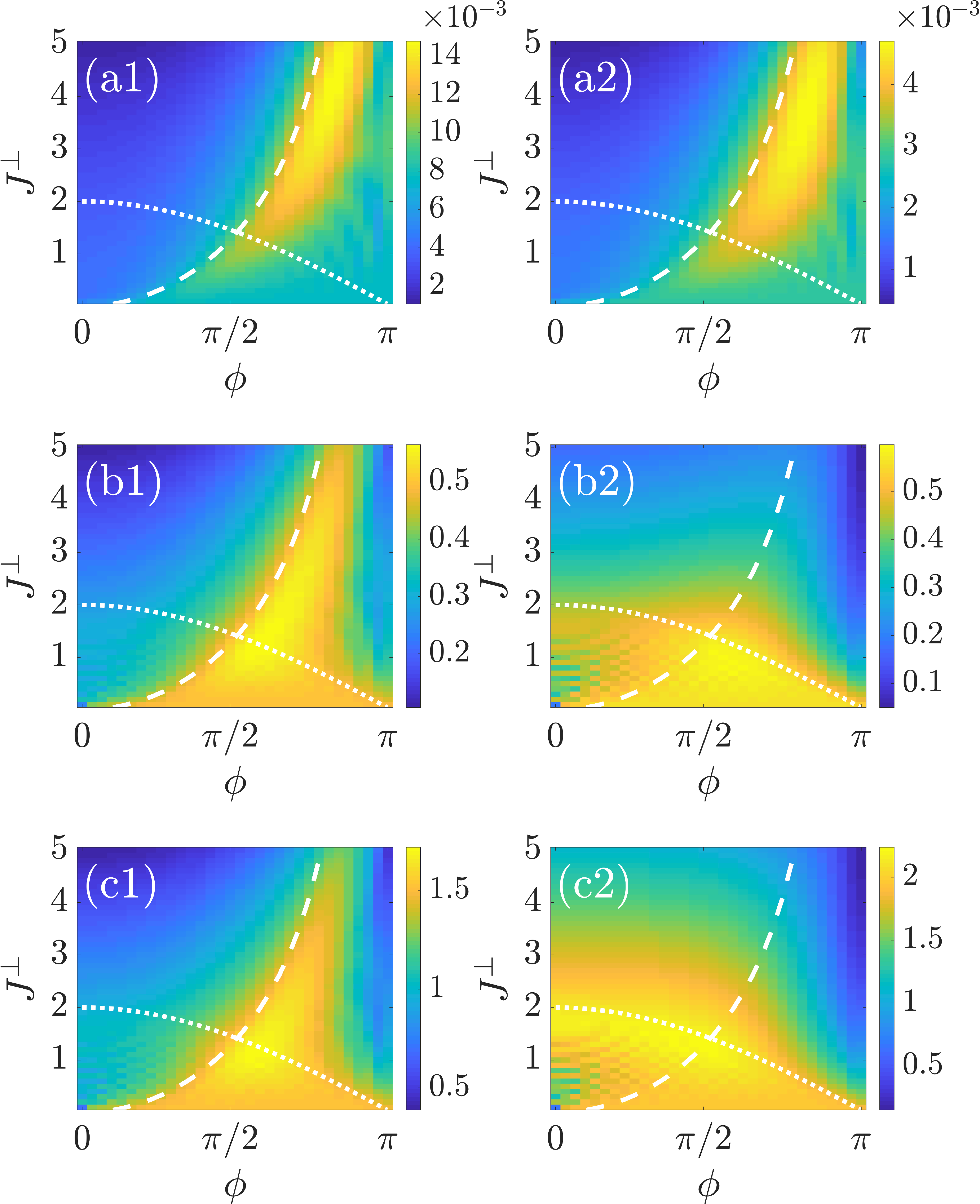}
    \caption{Total particle current $\mathcal{J}_{P}$, panels (a1), (b1) and (c1), and heat current $\mathcal{J}_{Q}$, panels (a2), (b2) and (c2), as a function of $J^{\perp}$ and $\phi$. 
    The bath parameters are $\gamma=0.1$, $T_{\mathrm{L}}=0.1$ and $T_{\mathrm{R}}=0.2$, (a1) and (a2), $T_{\mathrm{R}}=2.0$, (b1) and (b2), and $T_{\mathrm{R}}=5.0$, (c1) and (c2). The left and right bath chemical potential is $\mu=E_{0}-\Delta$ with $\Delta=0.1$ and the length of the ladder is $L=32$.
 }
    \label{fig:phcurr01}
\end{figure} 

The temperature dependence of the currents in the ladder is shown in Fig. \ref{fig:phcurr01}. First, we analyse the scenario for which the system-bath coupling strength $\gamma=0.1$. The top panels correspond to the low temperature bias where the particle and heat current have a similar pattern. This is because when the temperature bias is small and the temperature of the baths are low, most particles transported have energy close to the ground state.

As the temperature bias is increased, the particle and heat currents exhibit different responses for the different regions (I)-(IV) presented in Fig. \ref{fig:espec}. In particular, we observe that in region (IV) a relatively large $\mathcal{J}_{P}$ is accompanied by a relatively weak $\mathcal{J}_{Q}$ and vice versa in region (II). This discrepancy can be understood in terms of the energy band structure. In region (IV), the two energy bands are narrow and are separated by a gap. At a moderate temperature (Fig. \ref{fig:phcurr01}(b1-b2)) the upper band is inaccessible because of the gap. Most of the particles transported are in the lower narrow band which has energy close to the ground state. Hence, a large particle current and a low value of heat current are obtained in this regime. For region (II), the energy bands are gapless and wide. The absence of gap results in the occupation of upper bands which are higher in energy. Although the occupation of the upper bands is small compared to the lower band, the particles in these bands have much larger energy compared to the ground-state ones. Hence, these particles contribute to stronger $\mathcal{J}_{Q}$.  Similarly, we see strong heat current in region (I) with a weak particle current when the temperature is increased to $T_{\mathrm{R}}=5.0$ as the upper band separated by a finite gap is populated. 

\begin{figure}[htp]
    \centering
    \includegraphics[width=\columnwidth]{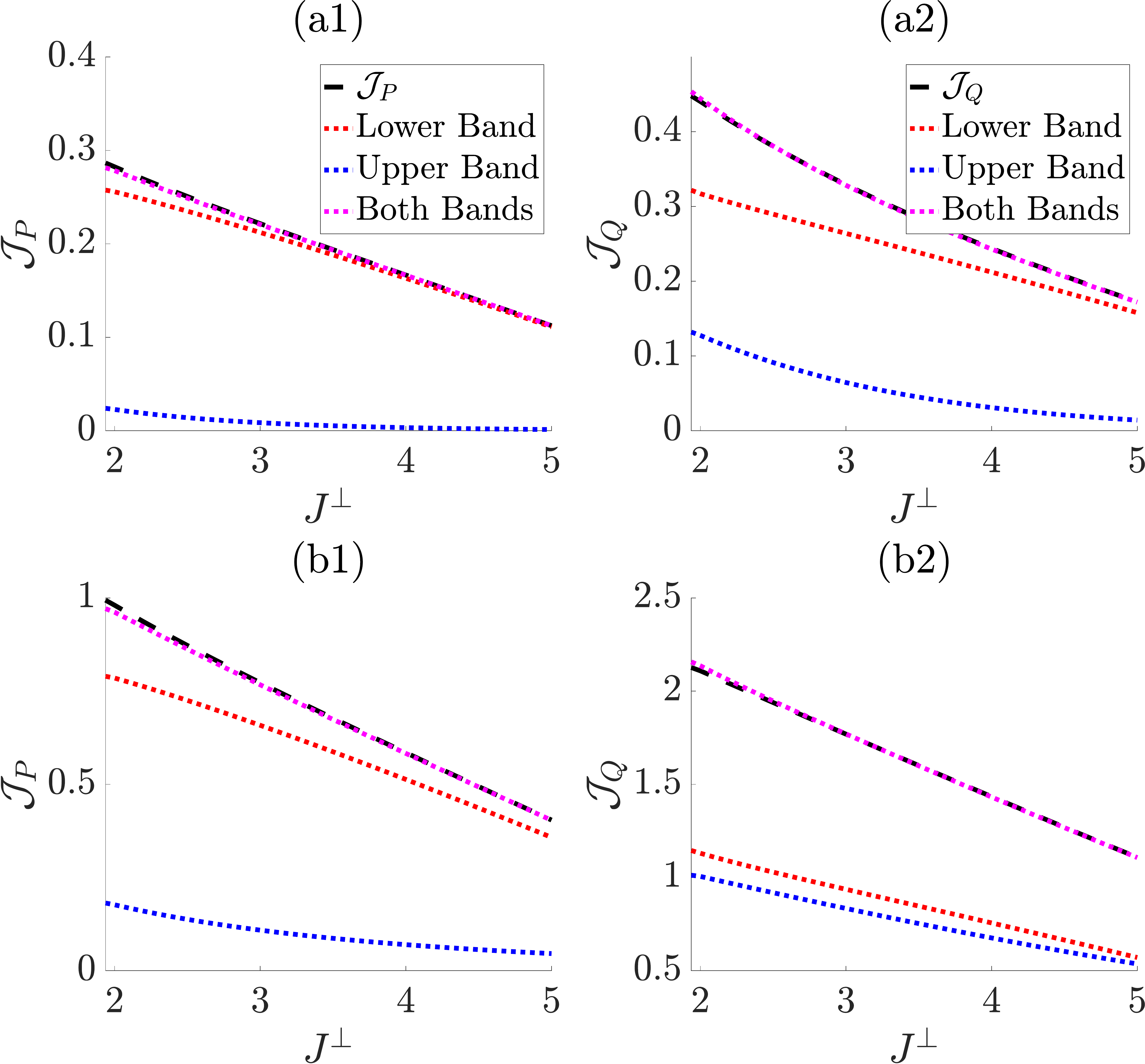}
    \caption{Approximated current contribution from the energy bands in region (I) with coupling $\gamma=0.1$.
    The left panels (a1) and (b1) correspond to particle current $\mathcal{J}_{P}$ and the right panels, (a2) and (b2), correspond to the heat current $\mathcal{J}_{Q}$. The top panels (a1) and (a2) are for $T_{\mathrm{R}}=2.0$ while the bottom panels (b1) and (b2) are for $T_{\mathrm{R}}=5.0$.  
    The red (blue) dotted line represents the current contribution from the lower (upper) band, and the magenta dotted line is the sum of the current contribution from the two bands.
    The dashed line shows the total current in the ladder obtained from Eq. (\ref{eqn:pmeirwingreen},\ref{eqn:hmeirwingreen}). The baths chemical potential is $\mu=E_{0}-\Delta$ with $\Delta=0.1$ and the length of the ladder is $L=32$. The other bath parameters are $T_{\mathrm{L}}=0.1$ and $\gamma=0.1$.}
    \label{fig:bandcurr}
\end{figure}

\begin{figure}[htp]
    \centering
    \includegraphics[width=\columnwidth]{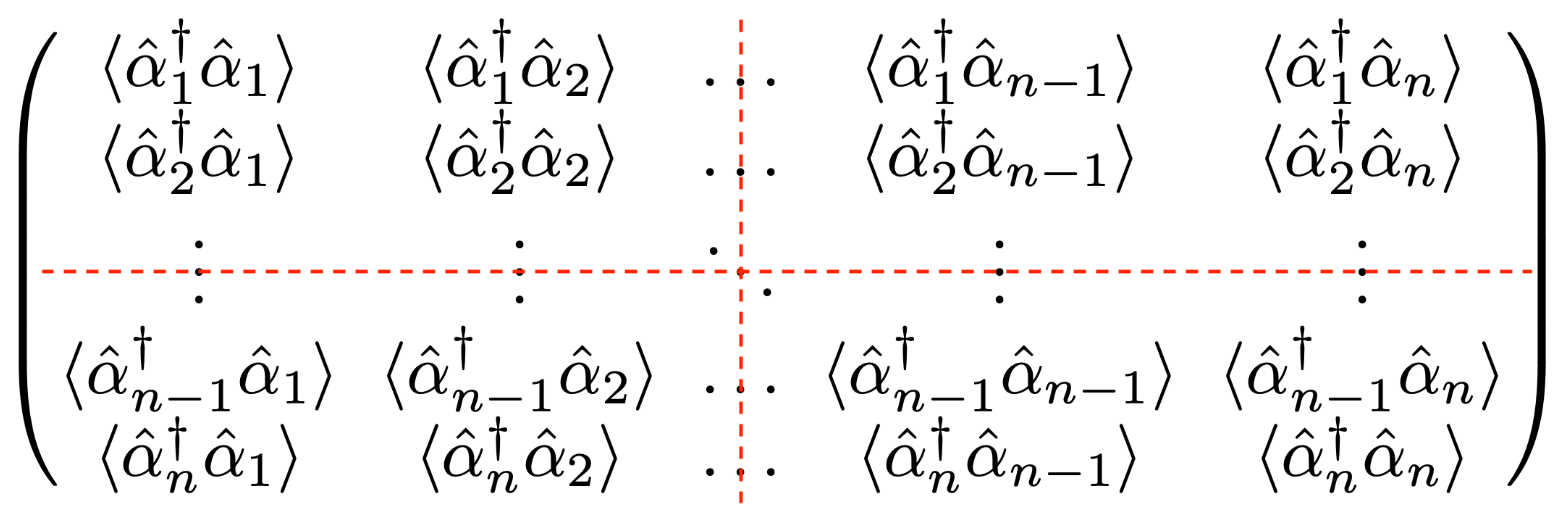}
    \caption{The single-particle density matrix in the single-particle energy eigenbasis.
    First diagonal quadrant: occupation and coherence within the lower band.
    Second diagonal quadrant: occupation and coherence within the upper band.
    Off-diagonal quadrants: coherence between states in different bands.
}
    \label{fig:denmatrix}
\end{figure}

To corroborate our understanding we further plot the $\mathcal{J}_{P}$ and $\mathcal{J}_{Q}$ contributions from each energy band for different values of $J^{\perp}$ when $\phi=0.5$ in Fig. \ref{fig:bandcurr}. Here, we focus on region (I) in which the energy spectrum contains a gap. In Fig. \ref{fig:bandcurr}(a1,b1) we depict the particle current $\mathcal{J}_P$, while in Fig. \ref{fig:bandcurr}(a2,b2) the heat current $\mathcal{J}_{Q}$. Furthermore, Fig. \ref{fig:bandcurr}(a1,a2) are for a smaller hot bath temperature $T_{\mathrm{R}}=2.0$, while Fig. \ref{fig:bandcurr}(b1,b2) have a larger one at $T_{\mathrm{R}}=5.0$.
In order to compute the contribution of the current from the different bands we divide the single-particle density matrix, written using the energy eigenbasis from the lower to the higher energy, in four sectors as depicted in Fig. \ref{fig:denmatrix}. The top left block includes the occupation and coherences for eigenstates in the lower band. The bottom right block includes occupation and coherence for eigenstates in the upper band. The two off-diagonal blocks describe the coherence between the two bands. One can thus compute the particle or heat current that would result from only considering the lower band (red dotted lines), only the upper band (blue dotted lines), or the sum of the two (magenta dotted lines) in Fig.\ref{fig:bandcurr}, while the total current (considering also interband coherence) is given by the black dashed line. In all panels, we observe that the interband coherence does not play a major role in these parameter regions when the energy gap is present. We also observe that only for the heat current $\mathcal{J}_Q$ and for high enough temperature, the current from the lower and the upper band is similar (see panel (b2)), despite the particle current in the lower band being larger than that in upper band (see panel (b1)). This is because the particles in the upper band carry more energy than in the lower band, and high temperatures allow the upper band to be partially populated.

\subsection{System-bath coupling strength $\gamma=0.5$}\label{ssec:g05}    
\begin{figure}[htp]
    \centering
    \includegraphics[width=\columnwidth]{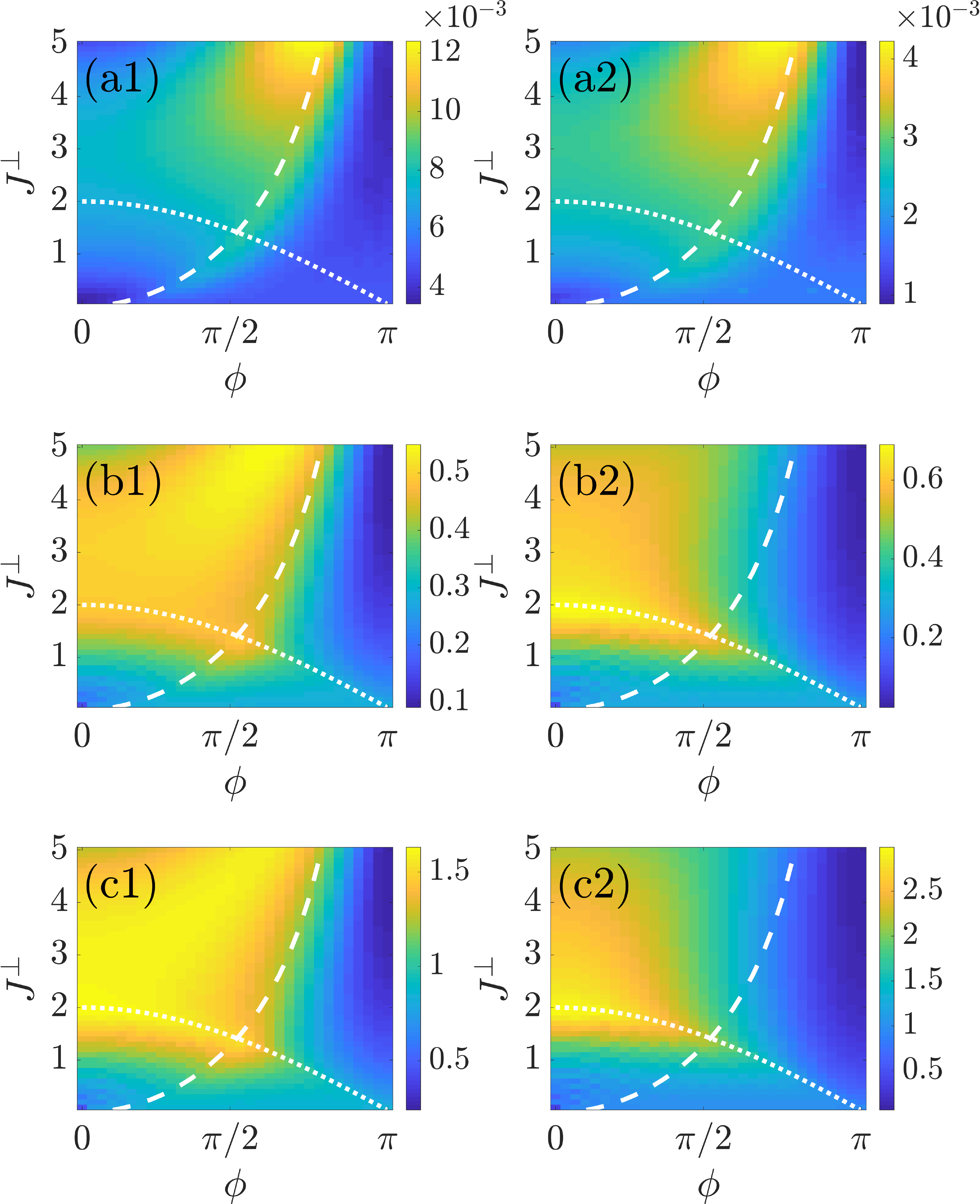}
    \caption{Identical analysis as for Fig. \ref{fig:phcurr01}, except with a stronger system-bath coupling $\gamma=0.5$.}    
    \label{fig:phcurr05}
\end{figure}

In Fig. \ref{fig:phcurr05}, we explore the transport through the system for a larger system-bath coupling $\gamma=0.5$. Fig. \ref{fig:phcurr05} shows that the pattern of $\mathcal{J}_{P}$ and $\mathcal{J}_{Q}$ at $\gamma=0.5$, is significantly different from $\gamma=0.1$. In particular, we observe that in regions (III) and (IV), which correspond to an underlying vortex phase, transport is much more suppressed compared to the case $\gamma=0.1$ in Fig. \ref{fig:phcurr01}. 
We also observe that the transport is favoured in region (I) where the presence of the energy gap in general hinders the transport.

\begin{figure}[htp]
    \centering
    \includegraphics[width=\columnwidth]{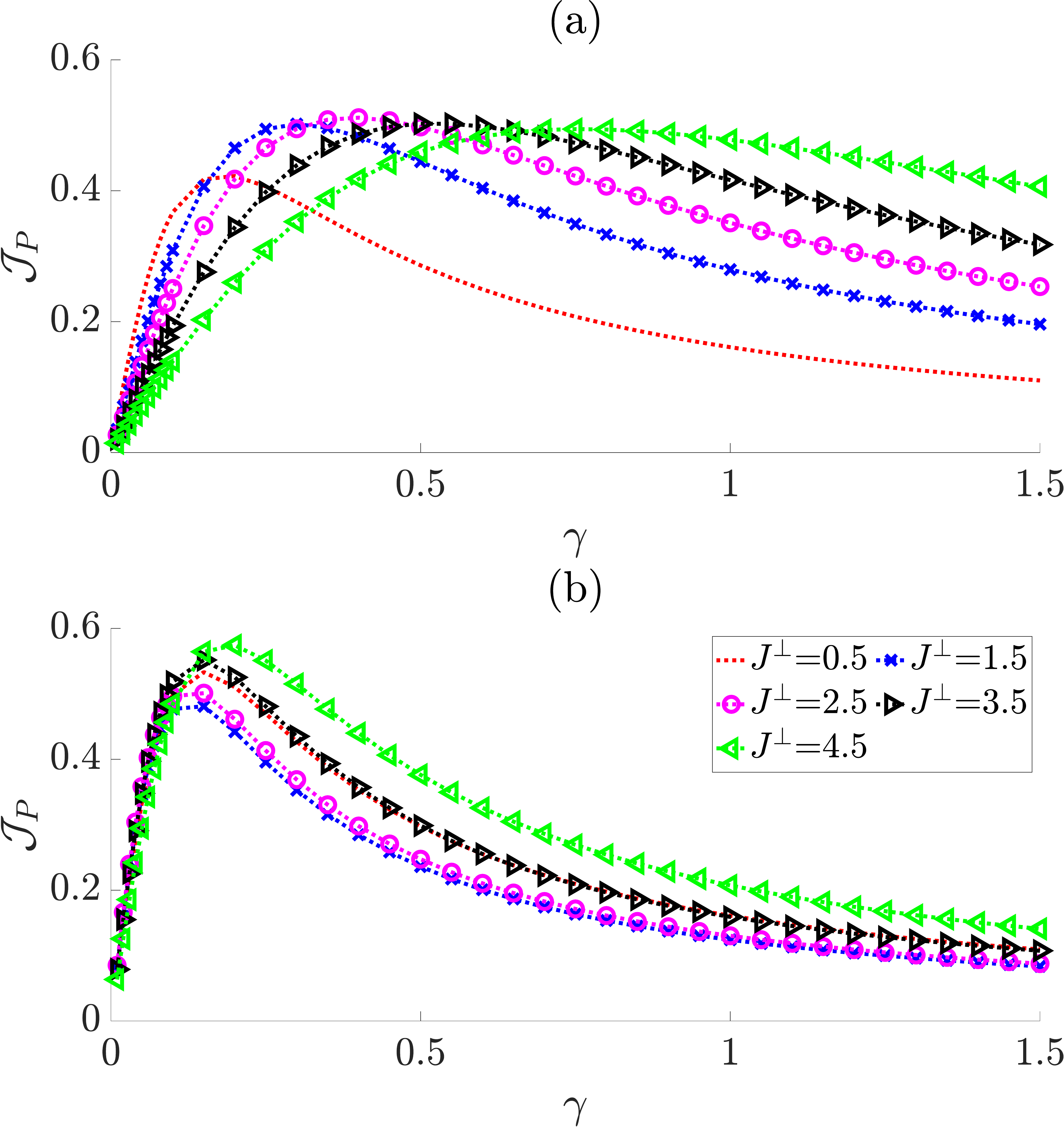}
    \caption{$\mathcal{J}_{P}$ versus $\gamma$ for different values of $J^{\perp}=0.5,1.5,2.5,3.5,4.5$.
    (a) $\phi=0.5$, for which the ground state is in the Meissner phase.
    (b) $\phi=2.5$, for which the ground state is in the vortex phase. The left and right bath chemical potential is $\mu=E_{0}-\Delta$ with $\Delta=0.1$ and the length of the ladder is $L=32$. 
    The other bath parameters are $T_{\mathrm{L}}=0.1$, $T_{\mathrm{R}}=2.0$.
}
    \label{fig:phasecurr}
\end{figure}

In order to understand the role of the system-bath coupling $\gamma$ in transport properties, we investigate the variation of $\mathcal{J}_{P}$  with $\gamma$ in Fig. \ref{fig:phasecurr}. The top panel corresponds to values of $J^{\perp}$ and $\phi$ for which the underlying ground state is in Meissner phase, regions (I) and (II). The bottom panel corresponds to the underlying vortex phase, regions (III) and (IV). In both scenarios, the current is non-monotonous, reaching a maximum for an intermediate value of the coupling $\gamma$, and tends to zero in the limit of extremely weak and strong coupling. This behaviour is expected as for small $\gamma$ the current would increase for larger interactions with the baths, but when the baths are too strongly coupled one faces quantum-Zeno-like dynamics \cite{MisraSudarshan1977, ChiuMisra1977, Kraus1981, ItanoWineland1990}.
However, the dependence of current on couplings is different in the two phases. In particular, we see that $\mathcal{J}_{P}$ is more robust to changes in $\gamma$ in the Meissner phase, especially at larger values of $J^{\perp}$, when compared to the vortex phase.
Similar observations are drawn from the plots of chiral current and local current in Fig. \ref{fig:chiral} and Fig. \ref{fig:loccurr2}, where we find that the vortex regime is more susceptible to changes in coupling strength.     

\begin{figure}[htp]
    \centering
    \includegraphics[width=\columnwidth]{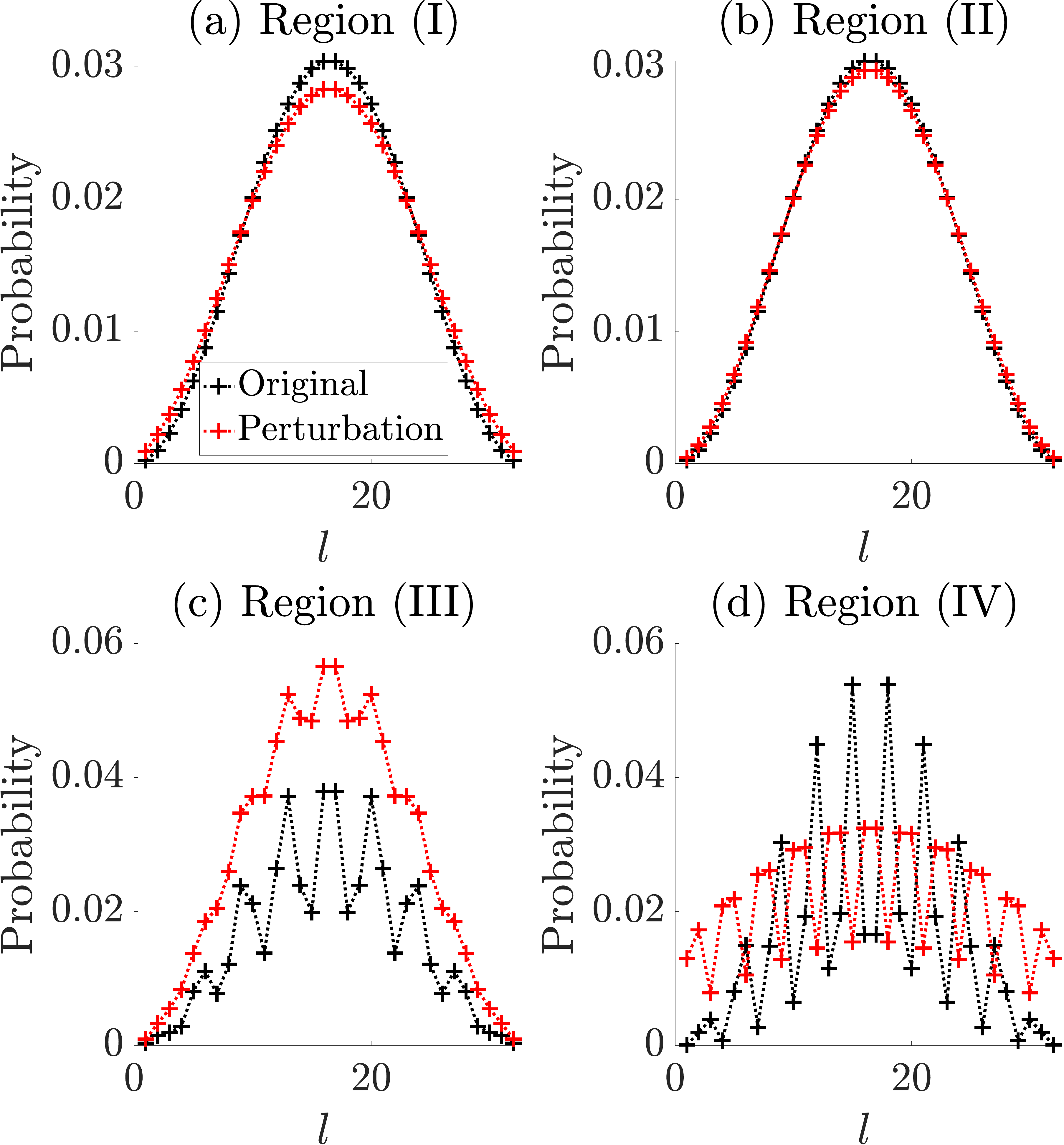}
    \caption{Probability distribution of the ground state of the system in the top leg of the ladder as a function of the site number $l$ for the four different regions shown in Fig. \ref{fig:chiral}.
    The amplitude for the bottom leg is not shown for clearer illustration.
    The black lines represent the ground state of the original unperturbed Hamiltonian $\hat{H}_{\rm S}$ given in Eq. (\ref{eqn:Hs}). The red lines represent the ground state of the perturbed system Hamiltonian $\hat{H}_{\rm pert}$.
}
    \label{fig:perturbation}
\end{figure}

To analyse this further, we consider the Hamiltonian of the ladder $\hat{H}_{\rm S}$ from Eq.(\ref{eqn:Hs}) and we introduce a small perturbation on the sites coupled to the baths to investigate the changes in the eigenstates of $\hat{H}_{\rm S}$. 
We consider an external perturbation given by $\hat{h}$
\begin{equation}\label{eqn:Hsp}
    \begin{aligned}
      \hat{h} =                      & - \xi \left[  J^{\|}\left( e^{i\phi/2} \hat{a}^{\dagger}_{1,1}\hat{a}_{2,1} + e^{i\phi/2} \hat{a}^{\dagger}_{L-1,1}\hat{a}_{L,1} \right) \right. \\
                      & \left. + J^{\perp} \left( \hat{a}^{\dagger}_{1,1}\hat{a}_{1,2} + \hat{a}^{\dagger}_{L,1}\hat{a}_{L,2} \right) + \text{H.c} \right]    
    ,\end{aligned}
\end{equation}
where $\xi$ is a small perturbation. The perturbed Hamiltonian $\hat{H}_{\rm pert} = \hat{H}_{\rm S} + \hat{h}$ is thus given by the ladder Hamiltonian $\hat{H}_{\rm S}$ plus an enhanced tunnelling terms which involve the sites that are coupled to the baths. 

In Fig. \ref{fig:perturbation}, we plot the ground state of $\hat{H}_{\rm S}$ and $\hat{H}_{\rm pert}$ in the four different regions for $\xi=0.1$. In regions (I) and (II), the ground states are in the Meissner phase. In regions (III) and (IV), the ground states are in the vortex phase. From Fig. \ref{fig:perturbation} we observe that the vortex phase is much more susceptible to the perturbations.
We thus expect that the properties of the systems in regions (III) and (IV) change more drastically under the influence of the baths. 

\section{Conclusions}\label{sec:conclusions}
To conclude, we have studied the interplay between boundary driving and gauge field using the minimal model of a non-interacting bosonic ladder coupled to two bosonic baths at its edges. The ground state of the bosonic ladder subjected to the artificial gauge field exhibits a quantum phase transition from the Meissner phase, with zero rung current, to the vortex phase which has finite rung current. The transition stems from the change in the geometry of the two-band energy structure of the bosonic ladder, where the number of minima increases from one to two. Using the non-equilibrium Green's function method, we study the robustness of this phase transition in the presence of bosonic baths. Our results show that the structure of the chiral currents are robust against the dissipative environment when the ground state of the ladder is in a Meissner phase. Despite the non-equilibrium set-up, we find maximal chiral current across the ground-state phase transition boundary. 
Meanwhile, the chiral current can be enhanced with increasing temperature bias. For moderate couplings, the remnants of the signatures of the underlying Meissner and vortex phases were observed, although at strong couplings the vortex-like behaviour is suppressed. 

We have also discussed particle and heat transport across the non-equilibrium ladder with temperature bias. When both baths are at low temperatures, we find that the steady-state phase diagram of particle and heat currents have similar patterns. When increasing the temperature bias, the particle and heat currents have different responses. In particular, we find a strong particle current with weak heat current and vice versa. We explain these intriguing behaviour in terms of the two-band energy structure of the bosonic ladder and the opening of a gap between the two bands. Furthermore, we demonstrate a strong dependence of system-bath coupling $\gamma$ on transport. 
Finally, using a perturbed Hamiltonian we show that the ladder is more resilient to coupling to the bath when the ground state is in the Meissner phase as compared to the case when the ground state is in the vortex phase.

Our study has shown that the energy band structure in the ladder can be tuned to steer and control the particle, heat, and also chiral currents. In general, the introduction of on-site interactions may significantly alter the transport properties of the driven system, as shown in Ref. \cite{GuoPoletti2017} for very different baths. It would thus be interesting to investigate, in the future, the role of on-site interactions. In addition, the effect of various system-bath coupling geometry is another direction to explore.

\section{Acknowledgement}
We acknowledge fruitful discussions with Jian-Sheng Wang and Hangbo Zhou. D.P. acknowledges supports from Singapore Ministry of Education, Singapore Academic Research Fund Tier-II (Project MOE2016-T2-1-065). The computational work for this article was partially performed on resources of the National Supercomputing Centre, Singapore (NSCC) \cite{NSCC}.   

\bibliography{ladder}

\begin{thebibliography}{66}%
\makeatletter
\providecommand \@ifxundefined [1]{%
 \@ifx{#1\undefined}
}%
\providecommand \@ifnum [1]{%
 \ifnum #1\expandafter \@firstoftwo
 \else \expandafter \@secondoftwo
 \fi
}%
\providecommand \@ifx [1]{%
 \ifx #1\expandafter \@firstoftwo
 \else \expandafter \@secondoftwo
 \fi
}%
\providecommand \natexlab [1]{#1}%
\providecommand \enquote  [1]{``#1''}%
\providecommand \bibnamefont  [1]{#1}%
\providecommand \bibfnamefont [1]{#1}%
\providecommand \citenamefont [1]{#1}%
\providecommand \href@noop [0]{\@secondoftwo}%
\providecommand \href [0]{\begingroup \@sanitize@url \@href}%
\providecommand \@href[1]{\@@startlink{#1}\@@href}%
\providecommand \@@href[1]{\endgroup#1\@@endlink}%
\providecommand \@sanitize@url [0]{\catcode `\\12\catcode `\$12\catcode
  `\&12\catcode `\#12\catcode `\^12\catcode `\_12\catcode `\%12\relax}%
\providecommand \@@startlink[1]{}%
\providecommand \@@endlink[0]{}%
\providecommand \url  [0]{\begingroup\@sanitize@url \@url }%
\providecommand \@url [1]{\endgroup\@href {#1}{\urlprefix }}%
\providecommand \urlprefix  [0]{URL }%
\providecommand \Eprint [0]{\href }%
\providecommand \doibase [0]{http://dx.doi.org/}%
\providecommand \selectlanguage [0]{\@gobble}%
\providecommand \bibinfo  [0]{\@secondoftwo}%
\providecommand \bibfield  [0]{\@secondoftwo}%
\providecommand \translation [1]{[#1]}%
\providecommand \BibitemOpen [0]{}%
\providecommand \bibitemStop [0]{}%
\providecommand \bibitemNoStop [0]{.\EOS\space}%
\providecommand \EOS [0]{\spacefactor3000\relax}%
\providecommand \BibitemShut  [1]{\csname bibitem#1\endcsname}%
\let\auto@bib@innerbib\@empty
\bibitem [{\citenamefont {Bertini}\ \emph {et~al.}(2020)\citenamefont
  {Bertini}, \citenamefont {{Heidrich-Meisner}}, \citenamefont {Karrasch},
  \citenamefont {Prosen}, \citenamefont {Steinigeweg},\ and\ \citenamefont
  {Znidaric}}]{BertiniZnidaric2020}%
  \BibitemOpen
  \bibfield  {author} {\bibinfo {author} {\bibfnamefont {B.}~\bibnamefont
  {Bertini}}, \bibinfo {author} {\bibfnamefont {F.}~\bibnamefont
  {{Heidrich-Meisner}}}, \bibinfo {author} {\bibfnamefont {C.}~\bibnamefont
  {Karrasch}}, \bibinfo {author} {\bibfnamefont {T.}~\bibnamefont {Prosen}},
  \bibinfo {author} {\bibfnamefont {R.}~\bibnamefont {Steinigeweg}}, \ and\
  \bibinfo {author} {\bibfnamefont {M.}~\bibnamefont {Znidaric}},\ }\href@noop
  {} {\bibfield  {journal} {\bibinfo  {journal} {arXiv:2003.03334}\ } (\bibinfo
  {year} {2020})}\BibitemShut {NoStop}%
\bibitem [{\citenamefont {Prosen}(2011)}]{Prosen2011b}%
  \BibitemOpen
  \bibfield  {author} {\bibinfo {author} {\bibfnamefont {T.}~\bibnamefont
  {Prosen}},\ }\href {\doibase 10.1103/PhysRevLett.106.217206} {\bibfield
  {journal} {\bibinfo  {journal} {Phys. Rev. Lett.}\ }\textbf {\bibinfo
  {volume} {106}},\ \bibinfo {pages} {217206} (\bibinfo {year}
  {2011})}\BibitemShut {NoStop}%
\bibitem [{\citenamefont {{\v Z}nidari{\v c}}(2011)}]{Znidaric2011}%
  \BibitemOpen
  \bibfield  {author} {\bibinfo {author} {\bibfnamefont {M.}~\bibnamefont {{\v
  Z}nidari{\v c}}},\ }\href {\doibase 10.1103/PhysRevLett.106.220601}
  {\bibfield  {journal} {\bibinfo  {journal} {Phys. Rev. Lett.}\ }\textbf
  {\bibinfo {volume} {106}},\ \bibinfo {pages} {220601} (\bibinfo {year}
  {2011})}\BibitemShut {NoStop}%
\bibitem [{\citenamefont {Benenti}\ \emph {et~al.}(2017)\citenamefont
  {Benenti}, \citenamefont {Casati}, \citenamefont {Saito},\ and\ \citenamefont
  {Whitney}}]{BenentiWhitney2017}%
  \BibitemOpen
  \bibfield  {author} {\bibinfo {author} {\bibfnamefont {G.}~\bibnamefont
  {Benenti}}, \bibinfo {author} {\bibfnamefont {G.}~\bibnamefont {Casati}},
  \bibinfo {author} {\bibfnamefont {K.}~\bibnamefont {Saito}}, \ and\ \bibinfo
  {author} {\bibfnamefont {R.~S.}\ \bibnamefont {Whitney}},\ }\href {\doibase
  10.1016/j.physrep.2017.05.008} {\bibfield  {journal} {\bibinfo  {journal}
  {Phys. Rep.}\ }\textbf {\bibinfo {volume} {694}},\ \bibinfo {pages} {1}
  (\bibinfo {year} {2017})}\BibitemShut {NoStop}%
\bibitem [{\citenamefont {Sachdev}(2011)}]{Sachdev2011}%
  \BibitemOpen
  \bibfield  {author} {\bibinfo {author} {\bibfnamefont {S.}~\bibnamefont
  {Sachdev}},\ }\href {\doibase 10.1017/CBO9780511973765} {\emph {\bibinfo
  {title} {Quantum {{Phase Transitions}}}}}\ (\bibinfo  {publisher} {{Cambridge
  University Press}},\ \bibinfo {address} {{Cambridge}},\ \bibinfo {year}
  {2011})\BibitemShut {NoStop}%
\bibitem [{\citenamefont {Diehl}\ \emph {et~al.}(2008)\citenamefont {Diehl},
  \citenamefont {Micheli}, \citenamefont {Kantian}, \citenamefont {Kraus},
  \citenamefont {B{\"u}chler},\ and\ \citenamefont {Zoller}}]{DiehlZoller2008}%
  \BibitemOpen
  \bibfield  {author} {\bibinfo {author} {\bibfnamefont {S.}~\bibnamefont
  {Diehl}}, \bibinfo {author} {\bibfnamefont {A.}~\bibnamefont {Micheli}},
  \bibinfo {author} {\bibfnamefont {A.}~\bibnamefont {Kantian}}, \bibinfo
  {author} {\bibfnamefont {B.}~\bibnamefont {Kraus}}, \bibinfo {author}
  {\bibfnamefont {H.~P.}\ \bibnamefont {B{\"u}chler}}, \ and\ \bibinfo {author}
  {\bibfnamefont {P.}~\bibnamefont {Zoller}},\ }\href {\doibase
  10.1038/nphys1073} {\bibfield  {journal} {\bibinfo  {journal} {Nat. Phys.}\
  }\textbf {\bibinfo {volume} {4}},\ \bibinfo {pages} {878} (\bibinfo {year}
  {2008})}\BibitemShut {NoStop}%
\bibitem [{\citenamefont {Diehl}\ \emph {et~al.}(2010)\citenamefont {Diehl},
  \citenamefont {Tomadin}, \citenamefont {Micheli}, \citenamefont {Fazio},\
  and\ \citenamefont {Zoller}}]{DiehlZoller2010}%
  \BibitemOpen
  \bibfield  {author} {\bibinfo {author} {\bibfnamefont {S.}~\bibnamefont
  {Diehl}}, \bibinfo {author} {\bibfnamefont {A.}~\bibnamefont {Tomadin}},
  \bibinfo {author} {\bibfnamefont {A.}~\bibnamefont {Micheli}}, \bibinfo
  {author} {\bibfnamefont {R.}~\bibnamefont {Fazio}}, \ and\ \bibinfo {author}
  {\bibfnamefont {P.}~\bibnamefont {Zoller}},\ }\href {\doibase
  10.1103/PhysRevLett.105.015702} {\bibfield  {journal} {\bibinfo  {journal}
  {Phys. Rev. Lett.}\ }\textbf {\bibinfo {volume} {105}},\ \bibinfo {pages}
  {015702} (\bibinfo {year} {2010})}\BibitemShut {NoStop}%
\bibitem [{\citenamefont {Dalla~Torre}\ \emph {et~al.}(2010)\citenamefont
  {Dalla~Torre}, \citenamefont {Demler}, \citenamefont {Giamarchi},\ and\
  \citenamefont {Altman}}]{DallaTorreAltman2010}%
  \BibitemOpen
  \bibfield  {author} {\bibinfo {author} {\bibfnamefont {E.~G.}\ \bibnamefont
  {Dalla~Torre}}, \bibinfo {author} {\bibfnamefont {E.}~\bibnamefont {Demler}},
  \bibinfo {author} {\bibfnamefont {T.}~\bibnamefont {Giamarchi}}, \ and\
  \bibinfo {author} {\bibfnamefont {E.}~\bibnamefont {Altman}},\ }\href
  {\doibase 10.1038/nphys1754} {\bibfield  {journal} {\bibinfo  {journal} {Nat.
  Phys.}\ }\textbf {\bibinfo {volume} {6}},\ \bibinfo {pages} {806} (\bibinfo
  {year} {2010})}\BibitemShut {NoStop}%
\bibitem [{\citenamefont {Ludwig}\ and\ \citenamefont
  {Marquardt}(2013)}]{LudwigMarquardt2013}%
  \BibitemOpen
  \bibfield  {author} {\bibinfo {author} {\bibfnamefont {M.}~\bibnamefont
  {Ludwig}}\ and\ \bibinfo {author} {\bibfnamefont {F.}~\bibnamefont
  {Marquardt}},\ }\href {\doibase 10.1103/PhysRevLett.111.073603} {\bibfield
  {journal} {\bibinfo  {journal} {Phys. Rev. Lett.}\ }\textbf {\bibinfo
  {volume} {111}},\ \bibinfo {pages} {073603} (\bibinfo {year}
  {2013})}\BibitemShut {NoStop}%
\bibitem [{\citenamefont {Carusotto}\ and\ \citenamefont
  {Ciuti}(2013)}]{CarusottoCiuti2013}%
  \BibitemOpen
  \bibfield  {author} {\bibinfo {author} {\bibfnamefont {I.}~\bibnamefont
  {Carusotto}}\ and\ \bibinfo {author} {\bibfnamefont {C.}~\bibnamefont
  {Ciuti}},\ }\href {\doibase 10.1103/RevModPhys.85.299} {\bibfield  {journal}
  {\bibinfo  {journal} {Rev. Mod. Phys.}\ }\textbf {\bibinfo {volume} {85}},\
  \bibinfo {pages} {299} (\bibinfo {year} {2013})}\BibitemShut {NoStop}%
\bibitem [{\citenamefont {Baumann}\ \emph {et~al.}(2010)\citenamefont
  {Baumann}, \citenamefont {Guerlin}, \citenamefont {Brennecke},\ and\
  \citenamefont {Esslinger}}]{BaumannEsslinger2010}%
  \BibitemOpen
  \bibfield  {author} {\bibinfo {author} {\bibfnamefont {K.}~\bibnamefont
  {Baumann}}, \bibinfo {author} {\bibfnamefont {C.}~\bibnamefont {Guerlin}},
  \bibinfo {author} {\bibfnamefont {F.}~\bibnamefont {Brennecke}}, \ and\
  \bibinfo {author} {\bibfnamefont {T.}~\bibnamefont {Esslinger}},\ }\href
  {\doibase 10.1038/nature09009} {\bibfield  {journal} {\bibinfo  {journal}
  {Nature}\ }\textbf {\bibinfo {volume} {464}},\ \bibinfo {pages} {1301}
  (\bibinfo {year} {2010})}\BibitemShut {NoStop}%
\bibitem [{\citenamefont {Kardar}(1986)}]{Kardar1986}%
  \BibitemOpen
  \bibfield  {author} {\bibinfo {author} {\bibfnamefont {M.}~\bibnamefont
  {Kardar}},\ }\href {\doibase 10.1103/PhysRevB.33.3125} {\bibfield  {journal}
  {\bibinfo  {journal} {Phys. Rev. B}\ }\textbf {\bibinfo {volume} {33}},\
  \bibinfo {pages} {3125} (\bibinfo {year} {1986})}\BibitemShut {NoStop}%
\bibitem [{\citenamefont {Granato}(1990)}]{Granato1990}%
  \BibitemOpen
  \bibfield  {author} {\bibinfo {author} {\bibfnamefont {E.}~\bibnamefont
  {Granato}},\ }\href {\doibase 10.1103/PhysRevB.42.4797} {\bibfield  {journal}
  {\bibinfo  {journal} {Phys. Rev. B}\ }\textbf {\bibinfo {volume} {42}},\
  \bibinfo {pages} {4797} (\bibinfo {year} {1990})}\BibitemShut {NoStop}%
\bibitem [{\citenamefont {Denniston}\ and\ \citenamefont
  {Tang}(1995)}]{DennistonTang1995}%
  \BibitemOpen
  \bibfield  {author} {\bibinfo {author} {\bibfnamefont {C.}~\bibnamefont
  {Denniston}}\ and\ \bibinfo {author} {\bibfnamefont {C.}~\bibnamefont
  {Tang}},\ }\href {\doibase 10.1103/PhysRevLett.75.3930} {\bibfield  {journal}
  {\bibinfo  {journal} {Phys. Rev. Lett.}\ }\textbf {\bibinfo {volume} {75}},\
  \bibinfo {pages} {3930} (\bibinfo {year} {1995})}\BibitemShut {NoStop}%
\bibitem [{\citenamefont {Nishiyama}(2000)}]{Nishiyama2000}%
  \BibitemOpen
  \bibfield  {author} {\bibinfo {author} {\bibfnamefont {Y.}~\bibnamefont
  {Nishiyama}},\ }\href {\doibase 10.1007/s100510070144} {\bibfield  {journal}
  {\bibinfo  {journal} {Eur. Phys. J. B}\ }\textbf {\bibinfo {volume} {17}},\
  \bibinfo {pages} {295} (\bibinfo {year} {2000})}\BibitemShut {NoStop}%
\bibitem [{\citenamefont {Orignac}\ and\ \citenamefont
  {Giamarchi}(2001)}]{OrignacGiamarchi2001}%
  \BibitemOpen
  \bibfield  {author} {\bibinfo {author} {\bibfnamefont {E.}~\bibnamefont
  {Orignac}}\ and\ \bibinfo {author} {\bibfnamefont {T.}~\bibnamefont
  {Giamarchi}},\ }\href {\doibase 10.1103/PhysRevB.64.144515} {\bibfield
  {journal} {\bibinfo  {journal} {Phys. Rev. B}\ }\textbf {\bibinfo {volume}
  {64}},\ \bibinfo {pages} {144515} (\bibinfo {year} {2001})}\BibitemShut
  {NoStop}%
\bibitem [{\citenamefont {Donohue}\ and\ \citenamefont
  {Giamarchi}(2001)}]{DonohueGiamarchi2001}%
  \BibitemOpen
  \bibfield  {author} {\bibinfo {author} {\bibfnamefont {P.}~\bibnamefont
  {Donohue}}\ and\ \bibinfo {author} {\bibfnamefont {T.}~\bibnamefont
  {Giamarchi}},\ }\href {\doibase 10.1103/PhysRevB.63.180508} {\bibfield
  {journal} {\bibinfo  {journal} {Phys. Rev. B}\ }\textbf {\bibinfo {volume}
  {63}},\ \bibinfo {pages} {180508} (\bibinfo {year} {2001})}\BibitemShut
  {NoStop}%
\bibitem [{\citenamefont {Dhar}\ \emph
  {et~al.}(2012{\natexlab{a}})\citenamefont {Dhar}, \citenamefont {Maji},
  \citenamefont {Mishra}, \citenamefont {Pai}, \citenamefont {Mukerjee},\ and\
  \citenamefont {Paramekanti}}]{DharParamekanti2012}%
  \BibitemOpen
  \bibfield  {author} {\bibinfo {author} {\bibfnamefont {A.}~\bibnamefont
  {Dhar}}, \bibinfo {author} {\bibfnamefont {M.}~\bibnamefont {Maji}}, \bibinfo
  {author} {\bibfnamefont {T.}~\bibnamefont {Mishra}}, \bibinfo {author}
  {\bibfnamefont {R.~V.}\ \bibnamefont {Pai}}, \bibinfo {author} {\bibfnamefont
  {S.}~\bibnamefont {Mukerjee}}, \ and\ \bibinfo {author} {\bibfnamefont
  {A.}~\bibnamefont {Paramekanti}},\ }\href {\doibase
  10.1103/PhysRevA.85.041602} {\bibfield  {journal} {\bibinfo  {journal} {Phys.
  Rev. A}\ }\textbf {\bibinfo {volume} {85}},\ \bibinfo {pages} {041602}
  (\bibinfo {year} {2012}{\natexlab{a}})}\BibitemShut {NoStop}%
\bibitem [{\citenamefont {Dhar}\ \emph {et~al.}(2013)\citenamefont {Dhar},
  \citenamefont {Mishra}, \citenamefont {Maji}, \citenamefont {Pai},
  \citenamefont {Mukerjee},\ and\ \citenamefont
  {Paramekanti}}]{DharParamekanti2013}%
  \BibitemOpen
  \bibfield  {author} {\bibinfo {author} {\bibfnamefont {A.}~\bibnamefont
  {Dhar}}, \bibinfo {author} {\bibfnamefont {T.}~\bibnamefont {Mishra}},
  \bibinfo {author} {\bibfnamefont {M.}~\bibnamefont {Maji}}, \bibinfo {author}
  {\bibfnamefont {R.~V.}\ \bibnamefont {Pai}}, \bibinfo {author} {\bibfnamefont
  {S.}~\bibnamefont {Mukerjee}}, \ and\ \bibinfo {author} {\bibfnamefont
  {A.}~\bibnamefont {Paramekanti}},\ }\href {\doibase
  10.1103/PhysRevB.87.174501} {\bibfield  {journal} {\bibinfo  {journal} {Phys.
  Rev. B}\ }\textbf {\bibinfo {volume} {87}},\ \bibinfo {pages} {174501}
  (\bibinfo {year} {2013})}\BibitemShut {NoStop}%
\bibitem [{\citenamefont {Petrescu}\ and\ \citenamefont
  {Le~Hur}(2013)}]{PetrescuLeHur2013}%
  \BibitemOpen
  \bibfield  {author} {\bibinfo {author} {\bibfnamefont {A.}~\bibnamefont
  {Petrescu}}\ and\ \bibinfo {author} {\bibfnamefont {K.}~\bibnamefont
  {Le~Hur}},\ }\href {\doibase 10.1103/PhysRevLett.111.150601} {\bibfield
  {journal} {\bibinfo  {journal} {Phys. Rev. Lett.}\ }\textbf {\bibinfo
  {volume} {111}},\ \bibinfo {pages} {150601} (\bibinfo {year}
  {2013})}\BibitemShut {NoStop}%
\bibitem [{\citenamefont {Cha}\ and\ \citenamefont {Shin}(2011)}]{ChaShin2011}%
  \BibitemOpen
  \bibfield  {author} {\bibinfo {author} {\bibfnamefont {M.-C.}\ \bibnamefont
  {Cha}}\ and\ \bibinfo {author} {\bibfnamefont {J.-G.}\ \bibnamefont {Shin}},\
  }\href {\doibase 10.1103/PhysRevA.83.055602} {\bibfield  {journal} {\bibinfo
  {journal} {Phys. Rev. A}\ }\textbf {\bibinfo {volume} {83}},\ \bibinfo
  {pages} {055602} (\bibinfo {year} {2011})}\BibitemShut {NoStop}%
\bibitem [{\citenamefont {Cr{\'e}pin}\ \emph {et~al.}(2011)\citenamefont
  {Cr{\'e}pin}, \citenamefont {Laflorencie}, \citenamefont {Roux},\ and\
  \citenamefont {Simon}}]{CrepinSimon2011}%
  \BibitemOpen
  \bibfield  {author} {\bibinfo {author} {\bibfnamefont {F.}~\bibnamefont
  {Cr{\'e}pin}}, \bibinfo {author} {\bibfnamefont {N.}~\bibnamefont
  {Laflorencie}}, \bibinfo {author} {\bibfnamefont {G.}~\bibnamefont {Roux}}, \
  and\ \bibinfo {author} {\bibfnamefont {P.}~\bibnamefont {Simon}},\ }\href
  {\doibase 10.1103/PhysRevB.84.054517} {\bibfield  {journal} {\bibinfo
  {journal} {Phys. Rev. B}\ }\textbf {\bibinfo {volume} {84}},\ \bibinfo
  {pages} {054517} (\bibinfo {year} {2011})}\BibitemShut {NoStop}%
\bibitem [{\citenamefont {Tovmasyan}\ \emph {et~al.}(2013)\citenamefont
  {Tovmasyan}, \citenamefont {{van Nieuwenburg}},\ and\ \citenamefont
  {Huber}}]{TovmasyanHuber2013}%
  \BibitemOpen
  \bibfield  {author} {\bibinfo {author} {\bibfnamefont {M.}~\bibnamefont
  {Tovmasyan}}, \bibinfo {author} {\bibfnamefont {E.~P.~L.}\ \bibnamefont {{van
  Nieuwenburg}}}, \ and\ \bibinfo {author} {\bibfnamefont {S.~D.}\ \bibnamefont
  {Huber}},\ }\href {\doibase 10.1103/PhysRevB.88.220510} {\bibfield  {journal}
  {\bibinfo  {journal} {Phys. Rev. B}\ }\textbf {\bibinfo {volume} {88}},\
  \bibinfo {pages} {220510} (\bibinfo {year} {2013})}\BibitemShut {NoStop}%
\bibitem [{\citenamefont {Wei}\ and\ \citenamefont
  {Mueller}(2014)}]{WeiMueller2014}%
  \BibitemOpen
  \bibfield  {author} {\bibinfo {author} {\bibfnamefont {R.}~\bibnamefont
  {Wei}}\ and\ \bibinfo {author} {\bibfnamefont {E.~J.}\ \bibnamefont
  {Mueller}},\ }\href {\doibase 10.1103/PhysRevA.89.063617} {\bibfield
  {journal} {\bibinfo  {journal} {Phys. Rev. A}\ }\textbf {\bibinfo {volume}
  {89}},\ \bibinfo {pages} {063617} (\bibinfo {year} {2014})}\BibitemShut
  {NoStop}%
\bibitem [{\citenamefont {Tokuno}\ and\ \citenamefont
  {Georges}(2014)}]{TokunoGeorges2014}%
  \BibitemOpen
  \bibfield  {author} {\bibinfo {author} {\bibfnamefont {A.}~\bibnamefont
  {Tokuno}}\ and\ \bibinfo {author} {\bibfnamefont {A.}~\bibnamefont
  {Georges}},\ }\href {\doibase 10.1088/1367-2630/16/7/073005} {\bibfield
  {journal} {\bibinfo  {journal} {New J. Phys.}\ }\textbf {\bibinfo {volume}
  {16}},\ \bibinfo {pages} {073005} (\bibinfo {year} {2014})}\BibitemShut
  {NoStop}%
\bibitem [{\citenamefont {Piraud}\ \emph {et~al.}(2015)\citenamefont {Piraud},
  \citenamefont {{Heidrich-Meisner}}, \citenamefont {McCulloch}, \citenamefont
  {Greschner}, \citenamefont {Vekua},\ and\ \citenamefont
  {Schollw{\"o}ck}}]{PiraudSchollwock2015}%
  \BibitemOpen
  \bibfield  {author} {\bibinfo {author} {\bibfnamefont {M.}~\bibnamefont
  {Piraud}}, \bibinfo {author} {\bibfnamefont {F.}~\bibnamefont
  {{Heidrich-Meisner}}}, \bibinfo {author} {\bibfnamefont {I.~P.}\ \bibnamefont
  {McCulloch}}, \bibinfo {author} {\bibfnamefont {S.}~\bibnamefont
  {Greschner}}, \bibinfo {author} {\bibfnamefont {T.}~\bibnamefont {Vekua}}, \
  and\ \bibinfo {author} {\bibfnamefont {U.}~\bibnamefont {Schollw{\"o}ck}},\
  }\href {\doibase 10.1103/PhysRevB.91.140406} {\bibfield  {journal} {\bibinfo
  {journal} {Phys. Rev. B}\ }\textbf {\bibinfo {volume} {91}},\ \bibinfo
  {pages} {140406} (\bibinfo {year} {2015})}\BibitemShut {NoStop}%
\bibitem [{\citenamefont {Greschner}\ \emph {et~al.}(2015)\citenamefont
  {Greschner}, \citenamefont {Piraud}, \citenamefont {{Heidrich-Meisner}},
  \citenamefont {McCulloch}, \citenamefont {Schollw{\"o}ck},\ and\
  \citenamefont {Vekua}}]{GreschnerVekua2015}%
  \BibitemOpen
  \bibfield  {author} {\bibinfo {author} {\bibfnamefont {S.}~\bibnamefont
  {Greschner}}, \bibinfo {author} {\bibfnamefont {M.}~\bibnamefont {Piraud}},
  \bibinfo {author} {\bibfnamefont {F.}~\bibnamefont {{Heidrich-Meisner}}},
  \bibinfo {author} {\bibfnamefont {I.~P.}\ \bibnamefont {McCulloch}}, \bibinfo
  {author} {\bibfnamefont {U.}~\bibnamefont {Schollw{\"o}ck}}, \ and\ \bibinfo
  {author} {\bibfnamefont {T.}~\bibnamefont {Vekua}},\ }\href {\doibase
  10.1103/PhysRevLett.115.190402} {\bibfield  {journal} {\bibinfo  {journal}
  {Phys. Rev. Lett.}\ }\textbf {\bibinfo {volume} {115}},\ \bibinfo {pages}
  {190402} (\bibinfo {year} {2015})}\BibitemShut {NoStop}%
\bibitem [{\citenamefont {Buser}\ \emph {et~al.}(2019)\citenamefont {Buser},
  \citenamefont {{Heidrich-Meisner}},\ and\ \citenamefont
  {Schollw{\"o}ck}}]{BuserSchollwock2019}%
  \BibitemOpen
  \bibfield  {author} {\bibinfo {author} {\bibfnamefont {M.}~\bibnamefont
  {Buser}}, \bibinfo {author} {\bibfnamefont {F.}~\bibnamefont
  {{Heidrich-Meisner}}}, \ and\ \bibinfo {author} {\bibfnamefont
  {U.}~\bibnamefont {Schollw{\"o}ck}},\ }\href {\doibase
  10.1103/PhysRevA.99.053601} {\bibfield  {journal} {\bibinfo  {journal} {Phys.
  Rev. A}\ }\textbf {\bibinfo {volume} {99}},\ \bibinfo {pages} {053601}
  (\bibinfo {year} {2019})}\BibitemShut {NoStop}%
\bibitem [{\citenamefont {Kamar}\ \emph {et~al.}(2019)\citenamefont {Kamar},
  \citenamefont {Kantian},\ and\ \citenamefont
  {Giamarchi}}]{KamarGiamarchi2019}%
  \BibitemOpen
  \bibfield  {author} {\bibinfo {author} {\bibfnamefont {N.~A.}\ \bibnamefont
  {Kamar}}, \bibinfo {author} {\bibfnamefont {A.}~\bibnamefont {Kantian}}, \
  and\ \bibinfo {author} {\bibfnamefont {T.}~\bibnamefont {Giamarchi}},\ }\href
  {\doibase 10.1103/PhysRevA.100.023614} {\bibfield  {journal} {\bibinfo
  {journal} {Phys. Rev. A}\ }\textbf {\bibinfo {volume} {100}},\ \bibinfo
  {pages} {023614} (\bibinfo {year} {2019})}\BibitemShut {NoStop}%
\bibitem [{\citenamefont {Fazio}\ and\ \citenamefont {{van der
  Zant}}(2001)}]{FaziovanderZant2001}%
  \BibitemOpen
  \bibfield  {author} {\bibinfo {author} {\bibfnamefont {R.}~\bibnamefont
  {Fazio}}\ and\ \bibinfo {author} {\bibfnamefont {H.}~\bibnamefont {{van der
  Zant}}},\ }\href {\doibase 10.1016/S0370-1573(01)00022-9} {\bibfield
  {journal} {\bibinfo  {journal} {Phys. Rep.}\ }\textbf {\bibinfo {volume}
  {355}},\ \bibinfo {pages} {235} (\bibinfo {year} {2001})}\BibitemShut
  {NoStop}%
\bibitem [{\citenamefont {{van Oudenaarden}}\ and\ \citenamefont
  {Mooij}(1996)}]{vanOudenaardenMooij1996}%
  \BibitemOpen
  \bibfield  {author} {\bibinfo {author} {\bibfnamefont {A.}~\bibnamefont {{van
  Oudenaarden}}}\ and\ \bibinfo {author} {\bibfnamefont {J.~E.}\ \bibnamefont
  {Mooij}},\ }\href {\doibase 10.1103/PhysRevLett.76.4947} {\bibfield
  {journal} {\bibinfo  {journal} {Phys. Rev. Lett.}\ }\textbf {\bibinfo
  {volume} {76}},\ \bibinfo {pages} {4947} (\bibinfo {year}
  {1996})}\BibitemShut {NoStop}%
\bibitem [{\citenamefont {Dalibard}\ \emph {et~al.}(2011)\citenamefont
  {Dalibard}, \citenamefont {Gerbier}, \citenamefont {Juzeli{\=u}nas},\ and\
  \citenamefont {{\"O}hberg}}]{DalibardOhberg2011}%
  \BibitemOpen
  \bibfield  {author} {\bibinfo {author} {\bibfnamefont {J.}~\bibnamefont
  {Dalibard}}, \bibinfo {author} {\bibfnamefont {F.}~\bibnamefont {Gerbier}},
  \bibinfo {author} {\bibfnamefont {G.}~\bibnamefont {Juzeli{\=u}nas}}, \ and\
  \bibinfo {author} {\bibfnamefont {P.}~\bibnamefont {{\"O}hberg}},\ }\href
  {\doibase 10.1103/RevModPhys.83.1523} {\bibfield  {journal} {\bibinfo
  {journal} {Rev. Mod. Phys.}\ }\textbf {\bibinfo {volume} {83}},\ \bibinfo
  {pages} {1523} (\bibinfo {year} {2011})}\BibitemShut {NoStop}%
\bibitem [{\citenamefont {Goldman}\ \emph {et~al.}(2014)\citenamefont
  {Goldman}, \citenamefont {Juzeli{\=u}nas}, \citenamefont {{\"O}hberg},\ and\
  \citenamefont {Spielman}}]{GoldmanSpielman2014}%
  \BibitemOpen
  \bibfield  {author} {\bibinfo {author} {\bibfnamefont {N.}~\bibnamefont
  {Goldman}}, \bibinfo {author} {\bibfnamefont {G.}~\bibnamefont
  {Juzeli{\=u}nas}}, \bibinfo {author} {\bibfnamefont {P.}~\bibnamefont
  {{\"O}hberg}}, \ and\ \bibinfo {author} {\bibfnamefont {I.~B.}\ \bibnamefont
  {Spielman}},\ }\href {\doibase 10.1088/0034-4885/77/12/126401} {\bibfield
  {journal} {\bibinfo  {journal} {Rep. Prog. Phys.}\ }\textbf {\bibinfo
  {volume} {77}},\ \bibinfo {pages} {126401} (\bibinfo {year}
  {2014})}\BibitemShut {NoStop}%
\bibitem [{\citenamefont {Atala}\ \emph {et~al.}(2014)\citenamefont {Atala},
  \citenamefont {Aidelsburger}, \citenamefont {Lohse}, \citenamefont
  {Barreiro}, \citenamefont {Paredes},\ and\ \citenamefont
  {Bloch}}]{AtalaBloch2014}%
  \BibitemOpen
  \bibfield  {author} {\bibinfo {author} {\bibfnamefont {M.}~\bibnamefont
  {Atala}}, \bibinfo {author} {\bibfnamefont {M.}~\bibnamefont {Aidelsburger}},
  \bibinfo {author} {\bibfnamefont {M.}~\bibnamefont {Lohse}}, \bibinfo
  {author} {\bibfnamefont {J.~T.}\ \bibnamefont {Barreiro}}, \bibinfo {author}
  {\bibfnamefont {B.}~\bibnamefont {Paredes}}, \ and\ \bibinfo {author}
  {\bibfnamefont {I.}~\bibnamefont {Bloch}},\ }\href {\doibase
  10.1038/nphys2998} {\bibfield  {journal} {\bibinfo  {journal} {Nat. Phys.}\
  }\textbf {\bibinfo {volume} {10}},\ \bibinfo {pages} {588} (\bibinfo {year}
  {2014})}\BibitemShut {NoStop}%
\bibitem [{\citenamefont {An}\ \emph {et~al.}(2017)\citenamefont {An},
  \citenamefont {Meier},\ and\ \citenamefont {Gadway}}]{AnGadway2017}%
  \BibitemOpen
  \bibfield  {author} {\bibinfo {author} {\bibfnamefont {F.~A.}\ \bibnamefont
  {An}}, \bibinfo {author} {\bibfnamefont {E.~J.}\ \bibnamefont {Meier}}, \
  and\ \bibinfo {author} {\bibfnamefont {B.}~\bibnamefont {Gadway}},\ }\href
  {\doibase 10.1126/sciadv.1602685} {\bibfield  {journal} {\bibinfo  {journal}
  {Sci. Adv.}\ }\textbf {\bibinfo {volume} {3}},\ \bibinfo {pages} {e1602685}
  (\bibinfo {year} {2017})}\BibitemShut {NoStop}%
\bibitem [{\citenamefont {Bermudez}\ \emph {et~al.}(2011)\citenamefont
  {Bermudez}, \citenamefont {Schaetz},\ and\ \citenamefont
  {Porras}}]{BermudezPorras2011}%
  \BibitemOpen
  \bibfield  {author} {\bibinfo {author} {\bibfnamefont {A.}~\bibnamefont
  {Bermudez}}, \bibinfo {author} {\bibfnamefont {T.}~\bibnamefont {Schaetz}}, \
  and\ \bibinfo {author} {\bibfnamefont {D.}~\bibnamefont {Porras}},\ }\href
  {\doibase 10.1103/PhysRevLett.107.150501} {\bibfield  {journal} {\bibinfo
  {journal} {Phys. Rev. Lett.}\ }\textbf {\bibinfo {volume} {107}},\ \bibinfo
  {pages} {150501} (\bibinfo {year} {2011})}\BibitemShut {NoStop}%
\bibitem [{\citenamefont {Guo}\ and\ \citenamefont
  {Poletti}(2016)}]{GuoPoletti2016}%
  \BibitemOpen
  \bibfield  {author} {\bibinfo {author} {\bibfnamefont {C.}~\bibnamefont
  {Guo}}\ and\ \bibinfo {author} {\bibfnamefont {D.}~\bibnamefont {Poletti}},\
  }\href {\doibase 10.1103/PhysRevA.94.033610} {\bibfield  {journal} {\bibinfo
  {journal} {Phys. Rev. A}\ }\textbf {\bibinfo {volume} {94}},\ \bibinfo
  {pages} {033610} (\bibinfo {year} {2016})}\BibitemShut {NoStop}%
\bibitem [{\citenamefont {Rivas}\ and\ \citenamefont
  {{Martin-Delgado}}(2017)}]{RivasMartin-Delgado2017}%
  \BibitemOpen
  \bibfield  {author} {\bibinfo {author} {\bibfnamefont {{\'A}.}~\bibnamefont
  {Rivas}}\ and\ \bibinfo {author} {\bibfnamefont {M.~A.}\ \bibnamefont
  {{Martin-Delgado}}},\ }\href {\doibase 10.1038/s41598-017-06722-x} {\bibfield
   {journal} {\bibinfo  {journal} {Sci. Rep.}\ }\textbf {\bibinfo {volume}
  {7}},\ \bibinfo {pages} {6350} (\bibinfo {year} {2017})}\BibitemShut
  {NoStop}%
\bibitem [{\citenamefont {Guo}\ and\ \citenamefont
  {Poletti}(2017)}]{GuoPoletti2017}%
  \BibitemOpen
  \bibfield  {author} {\bibinfo {author} {\bibfnamefont {C.}~\bibnamefont
  {Guo}}\ and\ \bibinfo {author} {\bibfnamefont {D.}~\bibnamefont {Poletti}},\
  }\href {\doibase 10.1103/PhysRevB.96.165409} {\bibfield  {journal} {\bibinfo
  {journal} {Phys. Rev. B}\ }\textbf {\bibinfo {volume} {96}},\ \bibinfo
  {pages} {165409} (\bibinfo {year} {2017})}\BibitemShut {NoStop}%
\bibitem [{\citenamefont {Gorini}\ \emph {et~al.}(1976)\citenamefont {Gorini},
  \citenamefont {Kossakowski},\ and\ \citenamefont
  {Sudarshan}}]{GoriniSudarshan1976}%
  \BibitemOpen
  \bibfield  {author} {\bibinfo {author} {\bibfnamefont {V.}~\bibnamefont
  {Gorini}}, \bibinfo {author} {\bibfnamefont {A.}~\bibnamefont {Kossakowski}},
  \ and\ \bibinfo {author} {\bibfnamefont {E.~C.~G.}\ \bibnamefont
  {Sudarshan}},\ }\href {\doibase 10.1063/1.522979} {\bibfield  {journal}
  {\bibinfo  {journal} {J. Math. Phys.}\ }\textbf {\bibinfo {volume} {17}},\
  \bibinfo {pages} {821} (\bibinfo {year} {1976})}\BibitemShut {NoStop}%
\bibitem [{\citenamefont {Lindblad}(1976)}]{Lindblad1976}%
  \BibitemOpen
  \bibfield  {author} {\bibinfo {author} {\bibfnamefont {G.}~\bibnamefont
  {Lindblad}},\ }\href {\doibase 10.1007/BF01608499} {\bibfield  {journal}
  {\bibinfo  {journal} {Commun. Math. Phys.}\ }\textbf {\bibinfo {volume}
  {48}},\ \bibinfo {pages} {119} (\bibinfo {year} {1976})}\BibitemShut
  {NoStop}%
\bibitem [{\citenamefont {Wichterich}\ \emph {et~al.}(2007)\citenamefont
  {Wichterich}, \citenamefont {Henrich}, \citenamefont {Breuer}, \citenamefont
  {Gemmer},\ and\ \citenamefont {Michel}}]{WichterichMichel2007}%
  \BibitemOpen
  \bibfield  {author} {\bibinfo {author} {\bibfnamefont {H.}~\bibnamefont
  {Wichterich}}, \bibinfo {author} {\bibfnamefont {M.~J.}\ \bibnamefont
  {Henrich}}, \bibinfo {author} {\bibfnamefont {H.-P.}\ \bibnamefont {Breuer}},
  \bibinfo {author} {\bibfnamefont {J.}~\bibnamefont {Gemmer}}, \ and\ \bibinfo
  {author} {\bibfnamefont {M.}~\bibnamefont {Michel}},\ }\href {\doibase
  10.1103/PhysRevE.76.031115} {\bibfield  {journal} {\bibinfo  {journal} {Phys.
  Rev. E}\ }\textbf {\bibinfo {volume} {76}},\ \bibinfo {pages} {031115}
  (\bibinfo {year} {2007})}\BibitemShut {NoStop}%
\bibitem [{\citenamefont {Thingna}\ \emph {et~al.}(2013)\citenamefont
  {Thingna}, \citenamefont {Wang},\ and\ \citenamefont
  {H{\"a}nggi}}]{ThingnaHanggi2013}%
  \BibitemOpen
  \bibfield  {author} {\bibinfo {author} {\bibfnamefont {J.}~\bibnamefont
  {Thingna}}, \bibinfo {author} {\bibfnamefont {J.-S.}\ \bibnamefont {Wang}}, \
  and\ \bibinfo {author} {\bibfnamefont {P.}~\bibnamefont {H{\"a}nggi}},\
  }\href {\doibase 10.1103/PhysRevE.88.052127} {\bibfield  {journal} {\bibinfo
  {journal} {Phys. Rev. E}\ }\textbf {\bibinfo {volume} {88}},\ \bibinfo
  {pages} {052127} (\bibinfo {year} {2013})}\BibitemShut {NoStop}%
\bibitem [{\citenamefont {Purkayastha}\ \emph {et~al.}(2016)\citenamefont
  {Purkayastha}, \citenamefont {Dhar},\ and\ \citenamefont
  {Kulkarni}}]{PurkayasthaKulkarni2016}%
  \BibitemOpen
  \bibfield  {author} {\bibinfo {author} {\bibfnamefont {A.}~\bibnamefont
  {Purkayastha}}, \bibinfo {author} {\bibfnamefont {A.}~\bibnamefont {Dhar}}, \
  and\ \bibinfo {author} {\bibfnamefont {M.}~\bibnamefont {Kulkarni}},\ }\href
  {\doibase 10.1103/PhysRevA.93.062114} {\bibfield  {journal} {\bibinfo
  {journal} {Phys. Rev. A}\ }\textbf {\bibinfo {volume} {93}},\ \bibinfo
  {pages} {062114} (\bibinfo {year} {2016})}\BibitemShut {NoStop}%
\bibitem [{\citenamefont {Xu}\ \emph {et~al.}(2017)\citenamefont {Xu},
  \citenamefont {Thingna},\ and\ \citenamefont {Wang}}]{XuWang2017}%
  \BibitemOpen
  \bibfield  {author} {\bibinfo {author} {\bibfnamefont {X.}~\bibnamefont
  {Xu}}, \bibinfo {author} {\bibfnamefont {J.}~\bibnamefont {Thingna}}, \ and\
  \bibinfo {author} {\bibfnamefont {J.-S.}\ \bibnamefont {Wang}},\ }\href
  {\doibase 10.1103/PhysRevB.95.035428} {\bibfield  {journal} {\bibinfo
  {journal} {Phys. Rev. B}\ }\textbf {\bibinfo {volume} {95}},\ \bibinfo
  {pages} {035428} (\bibinfo {year} {2017})}\BibitemShut {NoStop}%
\bibitem [{\citenamefont {Hofer}\ \emph {et~al.}(2017)\citenamefont {Hofer},
  \citenamefont {{Perarnau-Llobet}}, \citenamefont {Miranda}, \citenamefont
  {Haack}, \citenamefont {Silva}, \citenamefont {Brask},\ and\ \citenamefont
  {Brunner}}]{HoferBrunner2017}%
  \BibitemOpen
  \bibfield  {author} {\bibinfo {author} {\bibfnamefont {P.~P.}\ \bibnamefont
  {Hofer}}, \bibinfo {author} {\bibfnamefont {M.}~\bibnamefont
  {{Perarnau-Llobet}}}, \bibinfo {author} {\bibfnamefont {L.~D.~M.}\
  \bibnamefont {Miranda}}, \bibinfo {author} {\bibfnamefont {G.}~\bibnamefont
  {Haack}}, \bibinfo {author} {\bibfnamefont {R.}~\bibnamefont {Silva}},
  \bibinfo {author} {\bibfnamefont {J.~B.}\ \bibnamefont {Brask}}, \ and\
  \bibinfo {author} {\bibfnamefont {N.}~\bibnamefont {Brunner}},\ }\href
  {\doibase 10.1088/1367-2630/aa964f} {\bibfield  {journal} {\bibinfo
  {journal} {New J. Phys.}\ }\textbf {\bibinfo {volume} {19}},\ \bibinfo
  {pages} {123037} (\bibinfo {year} {2017})}\BibitemShut {NoStop}%
\bibitem [{\citenamefont {Caroli}\ \emph {et~al.}(1971)\citenamefont {Caroli},
  \citenamefont {Combescot}, \citenamefont {Nozieres},\ and\ \citenamefont
  {{Saint-James}}}]{CaroliSaint-James1971}%
  \BibitemOpen
  \bibfield  {author} {\bibinfo {author} {\bibfnamefont {C.}~\bibnamefont
  {Caroli}}, \bibinfo {author} {\bibfnamefont {R.}~\bibnamefont {Combescot}},
  \bibinfo {author} {\bibfnamefont {P.}~\bibnamefont {Nozieres}}, \ and\
  \bibinfo {author} {\bibfnamefont {D.}~\bibnamefont {{Saint-James}}},\ }\href
  {\doibase 10.1088/0022-3719/4/8/018} {\bibfield  {journal} {\bibinfo
  {journal} {J. Phys. C: Solid State Phys.}\ }\textbf {\bibinfo {volume} {4}},\
  \bibinfo {pages} {916} (\bibinfo {year} {1971})}\BibitemShut {NoStop}%
\bibitem [{\citenamefont {Meir}\ and\ \citenamefont
  {Wingreen}(1992)}]{MeirWingreen1992}%
  \BibitemOpen
  \bibfield  {author} {\bibinfo {author} {\bibfnamefont {Y.}~\bibnamefont
  {Meir}}\ and\ \bibinfo {author} {\bibfnamefont {N.~S.}\ \bibnamefont
  {Wingreen}},\ }\href {\doibase 10.1103/PhysRevLett.68.2512} {\bibfield
  {journal} {\bibinfo  {journal} {Phys. Rev. Lett.}\ }\textbf {\bibinfo
  {volume} {68}},\ \bibinfo {pages} {2512} (\bibinfo {year}
  {1992})}\BibitemShut {NoStop}%
\bibitem [{\citenamefont {Haug}\ and\ \citenamefont
  {Jauho}(2008)}]{HaugJauho2008}%
  \BibitemOpen
  \bibfield  {author} {\bibinfo {author} {\bibfnamefont {H.}~\bibnamefont
  {Haug}}\ and\ \bibinfo {author} {\bibfnamefont {A.-P.}\ \bibnamefont
  {Jauho}},\ }\href {\doibase 10.1007/978-3-540-73564-9} {\emph {\bibinfo
  {title} {Quantum {{Kinetics}} in {{Transport}} and {{Optics}} of
  {{Semiconductors}}}}}\ (\bibinfo  {publisher} {{Springer-Verlag}},\ \bibinfo
  {address} {{Berlin Heidelberg}},\ \bibinfo {year} {2008})\BibitemShut
  {NoStop}%
\bibitem [{\citenamefont {Prociuk}\ \emph {et~al.}(2010)\citenamefont
  {Prociuk}, \citenamefont {Phillips},\ and\ \citenamefont
  {Dunietz}}]{ProciukDunietz2010}%
  \BibitemOpen
  \bibfield  {author} {\bibinfo {author} {\bibfnamefont {A.}~\bibnamefont
  {Prociuk}}, \bibinfo {author} {\bibfnamefont {H.}~\bibnamefont {Phillips}}, \
  and\ \bibinfo {author} {\bibfnamefont {B.~D.}\ \bibnamefont {Dunietz}},\
  }\href {\doibase 10.1088/1742-6596/220/1/012008} {\bibfield  {journal}
  {\bibinfo  {journal} {J. Phys.: Conf. Ser.}\ }\textbf {\bibinfo {volume}
  {220}},\ \bibinfo {pages} {012008} (\bibinfo {year} {2010})}\BibitemShut
  {NoStop}%
\bibitem [{\citenamefont {Aeberhard}(2011)}]{Aeberhard2011}%
  \BibitemOpen
  \bibfield  {author} {\bibinfo {author} {\bibfnamefont {U.}~\bibnamefont
  {Aeberhard}},\ }\href {\doibase 10.1007/s10825-011-0375-6} {\bibfield
  {journal} {\bibinfo  {journal} {J. Comput. Electron.}\ }\textbf {\bibinfo
  {volume} {10}},\ \bibinfo {pages} {394} (\bibinfo {year} {2011})}\BibitemShut
  {NoStop}%
\bibitem [{\citenamefont {Zimbovskaya}\ and\ \citenamefont
  {Pederson}(2011)}]{ZimbovskayaPederson2011}%
  \BibitemOpen
  \bibfield  {author} {\bibinfo {author} {\bibfnamefont {N.~A.}\ \bibnamefont
  {Zimbovskaya}}\ and\ \bibinfo {author} {\bibfnamefont {M.~R.}\ \bibnamefont
  {Pederson}},\ }\href {\doibase 10.1016/j.physrep.2011.08.002} {\bibfield
  {journal} {\bibinfo  {journal} {Phys. Rep.}\ }\textbf {\bibinfo {volume}
  {509}},\ \bibinfo {pages} {1} (\bibinfo {year} {2011})}\BibitemShut {NoStop}%
\bibitem [{\citenamefont {Nikoli{\'c}}\ \emph {et~al.}(2012)\citenamefont
  {Nikoli{\'c}}, \citenamefont {Saha}, \citenamefont {Markussen},\ and\
  \citenamefont {Thygesen}}]{NikolicThygesen2012}%
  \BibitemOpen
  \bibfield  {author} {\bibinfo {author} {\bibfnamefont {B.~K.}\ \bibnamefont
  {Nikoli{\'c}}}, \bibinfo {author} {\bibfnamefont {K.~K.}\ \bibnamefont
  {Saha}}, \bibinfo {author} {\bibfnamefont {T.}~\bibnamefont {Markussen}}, \
  and\ \bibinfo {author} {\bibfnamefont {K.~S.}\ \bibnamefont {Thygesen}},\
  }\href {\doibase 10.1007/s10825-012-0386-y} {\bibfield  {journal} {\bibinfo
  {journal} {J. Comput. Electron.}\ }\textbf {\bibinfo {volume} {11}},\
  \bibinfo {pages} {78} (\bibinfo {year} {2012})}\BibitemShut {NoStop}%
\bibitem [{\citenamefont {Dhar}\ \emph
  {et~al.}(2012{\natexlab{b}})\citenamefont {Dhar}, \citenamefont {Saito},\
  and\ \citenamefont {H{\"a}nggi}}]{DharHanggi2012}%
  \BibitemOpen
  \bibfield  {author} {\bibinfo {author} {\bibfnamefont {A.}~\bibnamefont
  {Dhar}}, \bibinfo {author} {\bibfnamefont {K.}~\bibnamefont {Saito}}, \ and\
  \bibinfo {author} {\bibfnamefont {P.}~\bibnamefont {H{\"a}nggi}},\ }\href
  {\doibase 10.1103/PhysRevE.85.011126} {\bibfield  {journal} {\bibinfo
  {journal} {Phys. Rev. E}\ }\textbf {\bibinfo {volume} {85}},\ \bibinfo
  {pages} {011126} (\bibinfo {year} {2012}{\natexlab{b}})}\BibitemShut
  {NoStop}%
\bibitem [{\citenamefont {Wang}\ \emph {et~al.}(2014)\citenamefont {Wang},
  \citenamefont {Agarwalla}, \citenamefont {Li},\ and\ \citenamefont
  {Thingna}}]{WangThingna2014}%
  \BibitemOpen
  \bibfield  {author} {\bibinfo {author} {\bibfnamefont {J.-S.}\ \bibnamefont
  {Wang}}, \bibinfo {author} {\bibfnamefont {B.~K.}\ \bibnamefont {Agarwalla}},
  \bibinfo {author} {\bibfnamefont {H.}~\bibnamefont {Li}}, \ and\ \bibinfo
  {author} {\bibfnamefont {J.}~\bibnamefont {Thingna}},\ }\href {\doibase
  10.1007/s11467-013-0340-x} {\bibfield  {journal} {\bibinfo  {journal} {Front.
  Phys.}\ }\textbf {\bibinfo {volume} {9}},\ \bibinfo {pages} {673} (\bibinfo
  {year} {2014})}\BibitemShut {NoStop}%
\bibitem [{\citenamefont {Ryndyk}(2016)}]{Ryndyk2016}%
  \BibitemOpen
  \bibfield  {author} {\bibinfo {author} {\bibfnamefont {D.}~\bibnamefont
  {Ryndyk}},\ }\href {\doibase 10.1007/978-3-319-24088-6} {\emph {\bibinfo
  {title} {Theory of {{Quantum Transport}} at {{Nanoscale}}}}},\ Vol.\ \bibinfo
  {volume} {184}\ (\bibinfo  {publisher} {{Springer International
  Publishing}},\ \bibinfo {address} {{Cham}},\ \bibinfo {year}
  {2016})\BibitemShut {NoStop}%
\bibitem [{Note1()}]{Note1}%
  \BibitemOpen
  \bibinfo {note} {Simulations for longer systems, even $L=64$ and different
  local potentials $V/J^{\parallel }$ have also been performed and are
  consistent with the results here obtained.}\BibitemShut {Stop}%
\bibitem [{\citenamefont {Dittrich}\ \emph {et~al.}(1998)\citenamefont
  {Dittrich}, \citenamefont {H{\"a}nggi}, \citenamefont {Kramer}, \citenamefont
  {Sch{\"o}n}, \citenamefont {Ingold},\ and\ \citenamefont
  {Zwerger}}]{DittrichZwerger1998}%
  \BibitemOpen
  \bibfield  {author} {\bibinfo {author} {\bibfnamefont {T.}~\bibnamefont
  {Dittrich}}, \bibinfo {author} {\bibfnamefont {P.}~\bibnamefont
  {H{\"a}nggi}}, \bibinfo {author} {\bibfnamefont {B.}~\bibnamefont {Kramer}},
  \bibinfo {author} {\bibfnamefont {G.}~\bibnamefont {Sch{\"o}n}}, \bibinfo
  {author} {\bibfnamefont {G.-L.}\ \bibnamefont {Ingold}}, \ and\ \bibinfo
  {author} {\bibfnamefont {W.}~\bibnamefont {Zwerger}},\ }\href@noop {} {\emph
  {\bibinfo {title} {Quantum {{Transport}} and {{Dissipation}}}}}\ (\bibinfo
  {publisher} {{Wiley}},\ \bibinfo {year} {1998})\BibitemShut {NoStop}%
\bibitem [{\citenamefont {Landauer}(1957)}]{Landauer1957}%
  \BibitemOpen
  \bibfield  {author} {\bibinfo {author} {\bibfnamefont {R.}~\bibnamefont
  {Landauer}},\ }\href {\doibase 10.1147/rd.13.0223} {\bibfield  {journal}
  {\bibinfo  {journal} {IBM J. Res. Dev.}\ }\textbf {\bibinfo {volume} {1}},\
  \bibinfo {pages} {223} (\bibinfo {year} {1957})}\BibitemShut {NoStop}%
\bibitem [{\citenamefont {Landauer}(1970)}]{Landauer1970}%
  \BibitemOpen
  \bibfield  {author} {\bibinfo {author} {\bibfnamefont {R.}~\bibnamefont
  {Landauer}},\ }\href {\doibase 10.1080/14786437008238472} {\bibfield
  {journal} {\bibinfo  {journal} {Philos. Mag.}\ }\textbf {\bibinfo {volume}
  {21}},\ \bibinfo {pages} {863} (\bibinfo {year} {1970})}\BibitemShut
  {NoStop}%
\bibitem [{\citenamefont {Datta}(1995)}]{Datta1995}%
  \BibitemOpen
  \bibfield  {author} {\bibinfo {author} {\bibfnamefont {S.}~\bibnamefont
  {Datta}},\ }\href {\doibase 10.1017/CBO9780511805776} {\emph {\bibinfo
  {title} {Electronic {{Transport}} in {{Mesoscopic Systems}}}}}\ (\bibinfo
  {publisher} {{Cambridge University Press}},\ \bibinfo {address}
  {{Cambridge}},\ \bibinfo {year} {1995})\BibitemShut {NoStop}%
\bibitem [{\citenamefont {Misra}\ and\ \citenamefont
  {Sudarshan}(1977)}]{MisraSudarshan1977}%
  \BibitemOpen
  \bibfield  {author} {\bibinfo {author} {\bibfnamefont {B.}~\bibnamefont
  {Misra}}\ and\ \bibinfo {author} {\bibfnamefont {E.~C.~G.}\ \bibnamefont
  {Sudarshan}},\ }\href {\doibase 10.1063/1.523304} {\bibfield  {journal}
  {\bibinfo  {journal} {J. Math. Phys.}\ }\textbf {\bibinfo {volume} {18}},\
  \bibinfo {pages} {756} (\bibinfo {year} {1977})}\BibitemShut {NoStop}%
\bibitem [{\citenamefont {Chiu}\ \emph {et~al.}(1977)\citenamefont {Chiu},
  \citenamefont {Sudarshan},\ and\ \citenamefont {Misra}}]{ChiuMisra1977}%
  \BibitemOpen
  \bibfield  {author} {\bibinfo {author} {\bibfnamefont {C.~B.}\ \bibnamefont
  {Chiu}}, \bibinfo {author} {\bibfnamefont {E.~C.~G.}\ \bibnamefont
  {Sudarshan}}, \ and\ \bibinfo {author} {\bibfnamefont {B.}~\bibnamefont
  {Misra}},\ }\href {\doibase 10.1103/PhysRevD.16.520} {\bibfield  {journal}
  {\bibinfo  {journal} {Phys. Rev. D}\ }\textbf {\bibinfo {volume} {16}},\
  \bibinfo {pages} {520} (\bibinfo {year} {1977})}\BibitemShut {NoStop}%
\bibitem [{\citenamefont {Kraus}(1981)}]{Kraus1981}%
  \BibitemOpen
  \bibfield  {author} {\bibinfo {author} {\bibfnamefont {K.}~\bibnamefont
  {Kraus}},\ }\href {\doibase 10.1007/BF00726936} {\bibfield  {journal}
  {\bibinfo  {journal} {Found. Phys.}\ }\textbf {\bibinfo {volume} {11}},\
  \bibinfo {pages} {547} (\bibinfo {year} {1981})}\BibitemShut {NoStop}%
\bibitem [{\citenamefont {Itano}\ \emph {et~al.}(1990)\citenamefont {Itano},
  \citenamefont {Heinzen}, \citenamefont {Bollinger},\ and\ \citenamefont
  {Wineland}}]{ItanoWineland1990}%
  \BibitemOpen
  \bibfield  {author} {\bibinfo {author} {\bibfnamefont {W.~M.}\ \bibnamefont
  {Itano}}, \bibinfo {author} {\bibfnamefont {D.~J.}\ \bibnamefont {Heinzen}},
  \bibinfo {author} {\bibfnamefont {J.~J.}\ \bibnamefont {Bollinger}}, \ and\
  \bibinfo {author} {\bibfnamefont {D.~J.}\ \bibnamefont {Wineland}},\ }\href
  {\doibase 10.1103/PhysRevA.41.2295} {\bibfield  {journal} {\bibinfo
  {journal} {Phys. Rev. A}\ }\textbf {\bibinfo {volume} {41}},\ \bibinfo
  {pages} {2295} (\bibinfo {year} {1990})}\BibitemShut {NoStop}%
\bibitem [{NSC()}]{NSCC}%
  \BibitemOpen
  \href@noop {} {}\bibinfo {howpublished} {https://www.nscc.sg/}\BibitemShut
  {NoStop}%
\end{thebibliography}%

\end{document}